\documentclass[
apjs,
twocolappendix,
twocolumn
]{aastex631}

\usepackage[T1]{fontenc}
\usepackage{CJKutf8}
\usepackage{hyperref}
\usepackage{amsmath} 
\usepackage{physics}
\usepackage{amssymb}
\usepackage{enumerate}
\usepackage{bm}

\shorttitle{ Nuclear Network in {\tt Gmunu} }
\shortauthors{Cheong et al.}
\graphicspath{{./}{figures/}}
\begin{document}

\title{Toward First-Principles Multi-Messenger Predictions: Coupling Nuclear Networks with GR Radiation-MHD in {\tt Gmunu}}

\author[0000-0003-1449-3363]{Patrick Chi-Kit \surname{Cheong} \begin{CJK*}{UTF8}{bkai}(張志杰)\end{CJK*}}
\email{patrick.cheong@berkeley.edu}
\altaffiliation[]{N3AS Postdoctoral fellow}%Lines break automatically or can be forced with \\
\affiliation{Department of Physics, University of California, Berkeley, Berkeley, CA 94720, USA}
\affiliation{Center for Nonlinear Studies, Los Alamos National Laboratory, Los Alamos, NM 87545, USA}

\author[0000-0003-2624-0056]{Christopher L. Fryer}
\affiliation{Center for Nonlinear Studies, Los Alamos National Laboratory, Los Alamos, NM 87545, USA}



\begin{abstract}
We present a new implementation of nuclear reaction networks in the \texttt{G}eneral-relativistic \texttt{mu}ltigrid \texttt{nu}merical (\texttt{Gmunu}) code, a framework for general relativistic radiation magnetohydrodynamics (GRRMHD). 
The extended code self-consistently evolves nuclear species fully coupled to hydrodynamics, magnetic fields, and neutrino radiation transport under the conformal flatness approximation to Einstein's equations. 
Four approximate nuclear networks are incorporated, with stiff source terms integrated implicitly using implicit–explicit Runge–Kutta schemes. 
Validation is performed through a suite of benchmarks, including conserved-to-primitive recovery with a tabulated stellar equation of state, one-zone silicon burning, and hydrodynamic tests of shock tubes, acoustic pulses, and detonation fronts of Type~Ia supernovae. 
These tests confirm accurate coupling between nuclear reactions and fluid dynamics, conserving both electron and nuclear mass fractions to machine precision.

As an application, we perform spherically symmetric core-collapse supernova simulations. 
The models reproduce the expected non-exploding behavior of standard progenitors, while enhanced neutrino heating leads to shock revival. 
Including nuclear burning further alters the post-shock composition and dynamics, converting silicon and oxygen layers into iron-group nuclei and strengthening the explosion. 
This demonstrates the impact of explosive burning on both ejecta composition and shock evolution, and establishes the stability of the coupled GR radiation–MHD–nuclear framework. 
Although magnetic fields are not evolved in the present 1D application, the implementation is fully compatible with multidimensional GRMHD simulations.

This work represents the first GRRMHD code combining M1 neutrino transport with fully coupled nuclear burning, paving the way for multidimensional studies of supernovae and compact object mergers where nucleosynthesis feedback shapes multi-messenger signals.
\end{abstract}



\section{\label{sec:intro}Introduction}
Nuclear hydrodynamics plays a central role in many astrophysical explosion scenarios, ranging from type~Ia supernovae (SNe Ia), classical novae, and X-ray bursts---which are governed by thermonuclear runaway processes---to core-collapse supernovae (CCSNe) and compact object mergers, where nuclear reactions critically influence both the dynamics and observable signals. 
In these environments, nuclear burning not only powers electromagnetic emission but also drives the synthesis of new elements, thereby shaping the chemical evolution of galaxies. 
Recent multi-messenger observations, such as gravitational waves from binary neutron star mergers~\citep{2017PhRvL.119p1101A} and high-energy neutrinos and photons from CCSNe, highlight the need for models that consistently incorporate nuclear physics, as these signatures depend sensitively on the interplay of hydrodynamics, neutrino transport, and nucleosynthesis.

Modeling these systems requires a framework that can capture the interplay of strong gravity, neutrino transport, magnetic fields, and nuclear reactions. 
In particular, CCSNe and neutron star mergers involve conditions extending above and below nuclear saturation density, where relativistic effects and microphysics become equally important. 
General-relativistic neutrino magnetohydrodynamics (GR$\nu$MHD) codes with coupled nuclear reaction networks are therefore highly desirable for a faithful description of such explosive events. 

Significant progress has been made in incorporating reduced nuclear burning into hydrodynamics codes~\citep{2014ApJ...782...91N, 2015ApJ...808L..21C, 2015MNRAS.454.1238L, 2016ApJ...818..123B, 2017ApJ...843....2H, 2017ApJ...842...13W, 2021ApJ...921..113S, 2020ApJS..248...11B, 2023ApJ...951..112N}, though these studies are restricted to Newtonian/special-relativistic frameworks.
In parallel, large reaction networks such as {\tt SkyNet}~\citep{2017ApJS..233...18L} and {\tt WinNet}~\citep{2023ApJS..268...66R} are often employed in post-processing to calculate detailed nucleosynthesis from tracer particles.
While this strategy enables the use of thousands of isotopes, it cannot account for the dynamical feedback of nuclear energy release during the explosion.
Fully general-relativistic approaches that self-consistently combine energy-dependent radiation transport, magnetic fields, and nuclear burning remain rare.
Some progress has been achieved in general relativity~\citep{2012ApJ...749...37M, 2017PhRvD..96h3016U, 2019ApJ...870...98U, 2025ApJ...981..119F}, but most existing implementations rely on highly simplified reaction networks (e.g., CNO cycles) or lack simultaneous support for nuclear equations of state and neutrino transport, thereby constraining their applicability across diverse astrophysical scenarios.
The present work therefore represents not only an initial step toward a more comprehensive and physically complete treatment of stellar explosions, but also a bridge between large post-processing networks and reduced in-situ networks within a fully relativistic framework—laying the foundation for future multi-messenger predictions in which nucleosynthesis feedback plays a central role.

In this work, we extend the {\tt Gmunu} code~\citep{2020CQGra..37n5015C, 2021MNRAS.508.2279C, 2023ApJS..267...38C} by implementing composition evolution, a realistic stellar equation of state, and approximate nuclear networks. 
Since nuclear burning introduces stiff source terms, we evolve the reactions implicitly. 
As in our previous work~\citep{2022ApJS..261...22C, 2023ApJS..267...38C}, implicit--explicit (IMEX) Runge--Kutta schemes (see, e.g., \citealt{ASCHER1997151, pareschi2005implicit, 2015JCoPh.286..172C}) are employed to maintain stability while retaining reasonable step sizes. 
We validate the implementation through a hierarchy of tests, beginning with the stellar equation of state, followed by one-zone nuclear burning, and culminating in Newtonian verification tests, before demonstrating astrophysical applications in CCSNe. 

The paper is organized as follows. 
Section~\ref{sec:method} outlines the methodology and implementation of the stellar equation of state and nuclear networks. 
Section~\ref{sec:tests} presents verification tests.
Section~\ref{sec:application_tests} demonstrates CCSNe simulations with nuclear burning. 
Finally, Section~\ref{sec:conclusions} summarizes our results and discusses future directions. 

Unless explicitly stated, we work in geometrized Heaviside-Lorentz units, for which the speed of light $c$, gravitational constant $G$, solar mass $M_{\odot}$, vacuum permittivity $\epsilon_0$ and the Boltzmann constant $k_{\rm{B}}$ are all equal to one ( $c=G=M_{\odot}=\epsilon_0 = k_{\text{B}} = 1$ ).
Greek indices, running from 0 to 3, are used for 4-quantities while the Roman indices, running from 1 to 3, are used for 3-quantities.

\section{\label{sec:method}Methods}

\subsection{\label{sec:hydro}Formulations}
{\tt Gmunu} adopts the $3+1$ reference-metric formalism \citep{2014PhRvD..89h4043M, 2020PhRvD.101j4007M, 2020PhRvD.102j4001B}.
In this formalism, the metric can be written as
\begin{equation}
	\begin{aligned}
	ds^2 = & g_{\mu\nu}dx^\mu dx^\nu \\
		= & -\alpha^2  dt^2 + \gamma_{ij} \left( dx^i + \beta^i dt \right) \left( dx^j + \beta^j dt \right),
	\end{aligned}
\end{equation}
where $\alpha$ is the lapse function, $\beta^i$ is the spacelike shift vector, and $\gamma_{ij}$ is the spatial metric.
We adopt a conformal decomposition of the spatial metric $\gamma_{ij}$ with the conformal factor $\psi$:
\begin{equation}
	\gamma_{ij} = \psi^4 \bar{\gamma}_{ij},
\end{equation}
where $\bar{\gamma}_{ij}$ is the conformally related metric.
This conformally related metric can be expressed as the sum of a background \emph{time-independent} reference metric $\hat{\gamma}_{ij}$ and deviations $h_{ij}^{\rm dev}$.
In our current implementation, the reference metric $\hat{\gamma}_{ij}$ is the flat spacetime metric of the chosen coordinate system (i.e., either Cartesian, cylindrical, or spherical coordinates).
Note that, in conformally flat approximations, the spacetime deviations vanish and the reference metric is the conformally related metric (i.e., $\bar{\gamma}_{ij}=\hat{\gamma}_{ij}$).

{\tt Gmunu} solves the general relativistic radiation magnetohydrodynamics.
The evolution equations for the system are derived from the local conservation of rest-mass and energy-momentum, and from the homogeneous Faraday's law.
Moreover, depending on applications, evolution of the electron fraction and neutrino number densities are also available.
Details of the implementation of these equations are documented in \citep{2020CQGra..37n5015C, 2021MNRAS.508.2279C, 2022ApJS..261...22C, 2023ApJS..267...38C, 2024ApJ...975..116C}.
Below, we focus on the additional equations for nuclear hydrodynamics.

The evolution of the mass fraction of the $l$-th species $X_l$ is given by \citep{2025ApJ...981..119F}:
\begin{equation}\label{eq:Xl}
	\frac{1}{\sqrt{-g}} \partial_\mu \left( \sqrt{-g} \rho X_l u^\mu \right) = A_l m_{\rm u} \dot{n}_{\rm nuc},
\end{equation}
where $g$ is the determinant of the spacetime metric, $\rho$ is the rest-mass density of the fluid, $u^\mu$ is the fluid four-velocity, and $m_{\rm u}$ is the atomic mass unit. 
$A_l$ and $\dot{n}_{\rm nuc}$ are the mass number and the change rate of the $l$-th species, respectively.

As in our previous work (e.g., \cite{2021MNRAS.508.2279C}), the evolution equations of the mass fraction \eqref{eq:Xl} can be expressed in the form:
\begin{equation}
	\partial_t \bm{q} + \frac{1}{\sqrt{\hat{\gamma}}}\partial_j\left[\sqrt{\hat{\gamma}} \bm{f}^j\right] = \bm{s}.
\end{equation}
Here, 
\begin{align}
	q_{DX_l} &\equiv \sqrt{{\gamma}/\hat{\gamma}} \left[ \rho X_l W  \right] ,\\
	\left(f_{DX_l}\right)^i  &\equiv \sqrt{{\gamma}/\hat{\gamma}} \left[  \rho X_l W \hat{v}^i \right] , \\
	s_{DX_l} &\equiv \sqrt{{\gamma}/\hat{\gamma}} \left[ \alpha A_l m_{\rm u} \dot{n}_{\rm nuc}  \right]  \label{eq:Xl_source} ,
\end{align}
where $\hat{v}^i \equiv \left( \alpha v^i - \beta^i \right)$, $v^i$ is the three-velocity measured by an Eulerian observer at rest in the current spatial 3-hypersurface, and $W \equiv 1/\sqrt{1-v^i v_i}$ is the Lorentz factor.

\subsection{\label{sec:eos}Equations of State}
Equations of state (EoSs) that depend on isotopic composition are necessary for nuclear hydrodynamical simulations.
In this work, the Helmholtz-type stellar EoS is provided by coupling to the EoS driver \texttt{helmeos}~\citep{2000ApJS..126..501T}.
In this EoS, the fluid is composed of ideal gas ions, radiation, and degenerate or relativistic electrons, in local thermodynamic equilibrium.
Additionally, we include the average mass excess per baryon contribution to the internal energy density:
\begin{equation}\label{eq:eps}
	\epsilon = \epsilon_{\rm th} \left(\rho, T, \left\{ X_l \right\} \right) + \frac{1}{m_{\rm u}} \sum_l \Delta m_l {Y_l},
\end{equation}
where $\epsilon_{\rm th} \left(\rho, T, \left\{ X_l \right\} \right)$ is the thermal energy density provided by \texttt{helmeos}, $Y_l = X_l / A_l$ is the molar abundance, and 
\begin{equation}
	\Delta m_l = m_l - A_l m_{\rm u}
\end{equation}
is the mass excess of the $l$-th species with atomic mass $m_l$.
This formulation allows the internal energy to remain unchanged during nuclear reactions, with the thermal energy adjusting accordingly when the composition changes \citep{2015ApJ...806..275P, 2025ApJ...981...63H}.

As noted in \cite{2010CQGra..27k4103O}, the zero point for the specific internal energy $\epsilon$ is not arbitrary, as it affects the energy-momentum tensor and, thus, spacetime evolution.
Rest-mass contributions must not be included in $\epsilon$, and the reference rest mass in the EoS must be consistent across regimes.

\texttt{Gmunu} adopts the conserved-to-primitive conversion scheme of \cite{2021PhRvD.103b3018K}, where the minimum specific enthalpy is required.
With the specific energy given by \eqref{eq:eps}, the minimum specific enthalpy is estimated as $h_{\min} \approx \min(m_l / A_l / m_{\rm u})$.

\subsubsection{\label{sec:eos_bridging}Transition to Nuclear Statistical Equilibrium region}

As most of our targeted systems involve neutron stars, a nuclear EoS that handles matter at high density (e.g. $\rho \gtrsim 10^8~{\rm g \cdot cm^{-3}}$) is needed.
Currently, {\tt Gmunu} adopts tabulated nuclear EoSs assuming nuclear statistical equilibrium (NSE) between protons, neutrons, alpha particles, and a representative heavy nucleus for these regions.

Connecting the nuclear and stellar EoSs smoothly is desired for applications that involve both high and low density regions.
The implementation of the transition between these regions is problem dependent.
For example, various approaches have been proposed in the context of core-collapse supernovae~\citep{2002A&A...396..361R, 2006A&A...447.1049B, 2015ApJ...806..275P, 2020ApJS..248...11B, 2023ApJ...951..112N}.

Here, we adopt an approach similar to the one reported in \cite{2015ApJ...806..275P, 2023ApJ...951..112N}.
Specifically, in regions where the rest-mass density is covered by both nuclear and stellar EoSs, we choose the temperature thresholds $T_{\rm hi} = 5.8~{\rm GK}$ and $T_{\rm lo} = 5~{\rm GK}$.
Nuclear EoS is used when $T \geq T_{\rm hi}$, while stellar EoS is used when $T < T_{\rm lo}$.
In the mixed region $T_{\rm lo} \leq T < T_{\rm hi}$, thermodynamical quantities are obtained via linear interpolation between the two EoSs.
Finally, when the density is outside the nuclear EoS table, the stellar EoS is used, and vice versa.

In the nuclear EoS region (i.e. $T \geq T_{\rm hi}$), only $Y_{\rm e}$ is required and $X_l$ is not needed.
In the stellar EoS region (i.e. $T < T_{\rm lo}$), the mass fractions $X_l$ are needed, and the electron fraction is overwritten as $Y_{\rm e} = \sum_l X_l Z_l/A_l$.
In the mixed region, both $Y_{\rm e}$ and $X_l$ must be consistent.
Here, we assume NSE compositions, where $X_l$ is given by an NSE solver (see Section~\ref{sec:NSE_solver}).
To minimize calls to the NSE solver, we evolve both $Y_{\rm e}$ and $X_l$ throughout the domain, monitoring the difference between the advected $Y_{\rm e}$ and $Y_{\rm e}(\{X_l\})$ in the mixed region.
If the absolute difference exceeds $10^{-3}$, the NSE solver overwrites $X_l$.

Usually, despite different physics considered in nuclear and stellar EoSs, the physical properties at densities $\rho \sim 10^7 - 10^8~{\rm g \cdot cm^{-3}}$ are similar~\citep{2002A&A...396..361R}.
The EoSs start to differ at $\rho \gtrsim 10^7~{\rm g \cdot cm^{-3}}$.
Moreover, the specific energy $\epsilon$ in the nuclear EoS is usually insensitive to temperature when $T \lesssim 1~{\rm MeV} \approx 11.6~{\rm GK}$.
Therefore, when solving for $T$ given $\epsilon$, a solution may exist in both EoSs for $\rho \gtrsim 10^7~{\rm g \cdot cm^{-3}}$ and $T \lesssim 10^{10}~{\rm K}$.
In the context of CCSNe, the temperature in high-density regions is usually high enough to assume NSE~\citep{2002A&A...396..361R, 2023ApJ...951..112N}.
In this work, we prioritize the solution available in the nuclear EoS, using the stellar EoS only when necessary.
We remind the reader that the implementation of EoS bridging is problem dependent; careful choice of EoSs and bridging methods is required.

\subsection{\label{sec:network}Nuclear Network}
Several approximate nuclear networks are implemented in \texttt{Gmunu}.
Specifically, 7-, 13-, 19-, and 21-species networks are available via \texttt{iso7}, \texttt{aprox13}, \texttt{aprox19}, and \texttt{aprox21} \citep{1999ApJS..124..241T, 2000ApJS..129..377T}.
Here, we focus on the standard 13-species $\alpha$-chain network:
$\rm ^{ 4}He$,
$\rm ^{12}C$,
$\rm ^{16}O$,
$\rm ^{20}Ne$,
$\rm ^{24}Mg$,
$\rm ^{28}Si$,
$\rm ^{32}S$,
$\rm ^{36}Ar$,
$\rm ^{40}Ca$,
$\rm ^{44}Ti$,
$\rm ^{48}Cr$,
$\rm ^{52}Fe$,
$\rm ^{56}Ni$.

\subsubsection{\label{sec:coupling_to_hydro}Coupling to Hydrodynamics}
The timescale of the change rate of certain isotopes in the source term~\eqref{eq:Xl_source} can be comparable to or even shorter than the hydrodynamical timescale.
In such cases, the reaction-induced source terms can be very stiff from a numerical point of view.
Applying explicit time integration would be inefficient due to extremely strict constraints on the time step, potentially requiring prohibitively small steps to maintain stability.
Therefore, nuclear reactions are often treated implicitly, for example using operator splitting (most notably Strang~\citep{doi:10.1137/0705041})~\citep{2025ApJ...981...63H}, first-order implicit Euler schemes~\citep{2017ApJS..233...18L, 2023ApJ...951..112N}, or higher-order methods based on spectral deferred correction~\citep{2019ApJ...886..105Z}.

In this work, we adopt implicit-explicit (IMEX) Runge-Kutta schemes~\citep{ASCHER1997151, pareschi2005implicit}, which have been applied and tested previously in \texttt{Gmunu} for resistive magnetohydrodynamics~\citep{2022ApJS..261...22C} and radiation transfer~\citep{2023ApJS..267...38C}.
The key advantage of IMEX schemes is that they allow the stiff source terms from nuclear reactions to be treated implicitly, while the non-stiff hydrodynamical fluxes are handled explicitly.
This approach provides stability for stiff reactions without sacrificing efficiency in the advection-dominated hydrodynamics.
For implementation details of IMEX in \texttt{Gmunu}, we refer readers to \cite{2022ApJS..261...22C}.

An implicit update of the mass fraction $\{X_l\}$ from timestep $\texttt{n}$ (denoted $\bm{q}^{\texttt{n}}$) to $\texttt{n+1}$ (denoted $\bm{q}^{\texttt{n+1}}$) can be expressed as
\begin{equation}
	\bm{q}^{\texttt{n+1}} = \bm{q}^{\texttt{n}} + \Delta t \, \bm{s}_{\rm stiff}(\bm{q}^{\texttt{n+1}}),
\end{equation}
where $\bm{s}_{\rm stiff}$ represents the stiff source terms, given by equation~\eqref{eq:Xl_source} in this case.
To obtain $\bm{q}^{\texttt{n+1}}$, we solve the non-linear system
\begin{equation}
	\bm{f}(\bm{q}) \equiv - \bm{q} + \bm{q}^{\texttt{n}} + \Delta t \, \bm{s}_{\rm stiff}(\bm{q}).
\end{equation}
During the implicit step, we assume that the spacetime metric and all primitive variables (e.g., $\rho$, $v^i$, $T$) remain unchanged except for the mass fractions $\{X_l\}$.
Thus, the source term can be expressed as
\begin{equation}
	s_{DX_l} = \sqrt{\gamma/\hat{\gamma}} \left[ \alpha \rho \dot{X}_l \right],
\end{equation}
where $\{\dot{X}_l\}$ are provided by the nuclear network modules.

Traditionally, nuclear reactions are treated as sets of first-order ordinary differential equations (ODEs) and solved with standard ODE integrators~\citep{1986A&A...162..103M, 1999JCoAM.109..321H, 1999ApJS..124..241T, 2025ApJ...981...63H}.
Here, we solve the non-linear system using a multidimensional Broyden method, falling back to the Newton-Raphson method if the Broyden method fails.
The Jacobian $\partial f_i / \partial q_j$ is computed analytically, with derivatives of the mass fraction change rates provided by the nuclear network modules.
The implementation of the Broyden solver and Jacobian computation follows \cite{press1996numerical}.

It is worth noting that the neutrino transfer module in \texttt{Gmunu} shares the same IMEX infrastructure as nuclear burning.
However, in the current implementation, they operate in different density and temperature regimes, allowing both modules to be used simultaneously without conflict.

Shocks are typically under-resolved in astrophysical simulations due to the finite numerical resolution. 
To mitigate spurious results, it is common to suppress nuclear burning inside numerical shocks~\citep{fryxell1989hydrodynamics}, ensuring that burning only occurs once physically appropriate post-shock conditions are established. 
In our implementation, we adopt a simplified version of the approach described in \cite{2024ApJ...966..150Z, 2025RNAAS...9..113B}, tagging a cell as shocked when
\begin{equation} \label{eq:shock_condition}
	\begin{aligned}
		& \frac{\left\vert \nabla P \cdot \vec{v} \right\vert}{P \left\vert v \right\vert} > f_{\rm shock}, \\
		& \nabla \cdot \vec{v} < 0,
	\end{aligned}
\end{equation}
where $f_{\rm shock} = 2/3$ by default. 
Here, $\nabla P$ and $\nabla \cdot \vec{v}$ are computed using slope-limited reconstructions (e.g., TVD or PPM limiters) to suppress oscillations near discontinuities and return a monotonic, cell-centred slope.  
This criterion effectively identifies cells with significant pressure jumps along the flow direction that are also undergoing compression, marking them as part of the shock region.  
Although we have not conducted an extensive parameter study ourselves, previous investigations indicate that variations of $f_{\rm shock}$ within the range $0.5$--$1.0$ do not qualitatively alter the dynamics~\citep{2025RNAAS...9..113B}.  
Alternative formulations, such as normalizing the pressure gradient by the local grid spacing rather than projecting along the velocity, have been shown to perform equivalently~\citep{2020A&A...635A.169G}.  

For all cells flagged as shocked, nuclear burning is temporarily disabled to prevent unphysical energy generation. 
While this approach is simple, it provides a robust and computationally efficient method for shock treatment in nuclear hydrodynamics.  
Future work will include implementing a dedicated test to explicitly verify the impact of this prescription and exploring refinements such as adaptive shock-capturing schemes or burning suppression based on additional physical criteria.

\subsubsection{\label{sec:NSE_solver}NSE solver}
An NSE solver that computes isotope compositions under NSE for a given thermodynamic state is often required for consistent composition evolution.
For regions that reach NSE, applying a nuclear network is unnecessary: compositions there can be set to satisfy NSE conditions.
The NSE solver can also be used to initialize compositions at the start of a simulation.
Our implementation of the NSE solver closely follows the methods described in \cite{2007ApJ...656..313C, 2009ADNDT..95...96S, 2008ApJ...685L.129S}.
Below, we outline the implementation details.

The mass fractions in NSE are obtained by solving
\begin{equation} \label{eq:nse_eq}
	\begin{aligned}
	& \sum_l X_l \left( \mu^{\rm kin}_{\rm p}, \mu^{\rm kin}_{\rm n} \right) = 1, \\
	& \sum_l \frac{Z_l}{A_l} X_l\left( \mu^{\rm kin}_{\rm p}, \mu^{\rm kin}_{\rm n} \right) = Y_{\rm e},
	\end{aligned}
\end{equation}
numerically for the chemical potentials of the kinetic part of the free protons and neutrons $\mu^{\rm kin}_{\rm p}$ and $\mu^{\rm kin}_{\rm n}$.
Each mass fraction is
\begin{equation}
	\begin{aligned}
		X_l & \left( \mu^{\rm kin}_{\rm p}, \mu^{\rm kin}_{\rm n} \right) = \frac{m_l}{\rho} g_l \left( \frac{2 \pi m_l k_{\rm B}T}{h^2} \right)^{3/2} \\
		& \times \exp \left( \frac{Z_l \mu^{\rm kin}_{\rm p} + N_l \mu^{\rm kin}_{\rm n} + Q_l - \mu_l^{\rm C} + Z_l \mu_{\rm p}^{\rm C}}{k_{\rm B}T} \right),
	\end{aligned}
\end{equation}
with $N_l = A_l - Z_l$, $Q_l$ the nuclear binding energy, $h$ Planck's constant, $k_{\rm B}$ Boltzmann's constant. 
$g_l$ is the nuclear partition functions of nucleus $l$, which is in general temperature dependent.
Nevertheless, in this work, we simply set $g_l = 2$, and the inclusion of tabulating more realistic tables (e.g. \cite{2000ADNDT..75....1R}) is left as future work.
$\mu_l^{\rm C}$ is the Coulomb corrections, which is obtained by using the fitting form by \cite{1998PhRvE..58.4941C}:
\begin{equation}
	\begin{aligned}
		\frac{\mu_l^{\rm C}}{k_{\rm B} T} = & A_1 \left[ \sqrt{\Gamma_l \left( A_2 + \Gamma_l \right)} - A_2 \ln \left(\sqrt{\frac{\Gamma_l}{A_2}} + \sqrt{1+\frac{\Gamma_l}{A_2}} \right) \right] \\ 
		& + 2 A_3 \left[ \sqrt{\Gamma_l} - \arctan \left( \sqrt{\Gamma_l} \right)\right],
	\end{aligned}
\end{equation}
where $A_1 = -0.9052$, $A_2 = 0.6322$, and $A_3 = - \sqrt{3}/2 - A_1/\sqrt{A_2}$.
$\Gamma_l = Z_l^{5/3} \Gamma_{\rm e}$ is the ion-specific Coulomb coupling parameter with $\Gamma_{\rm e} = e^2 \left( 4 \pi n_{\rm e} / 3\right)^{1/3} / \left( k_{\rm B} T\right)$, where the electron number density is estimated as $n_{\rm e} = Y_{\rm e} \rho / m_{\rm u}$.

The constrained system~\eqref{eq:nse_eq} is solved numerically using the Newton-Raphson method.

\section{\label{sec:tests}Code tests}
Although our ultimate goal is to apply nuclear burning in general-relativistic simulations, it is essential to benchmark the EoS and nuclear-reaction modules individually.
Comparisons with well-established Newtonian test problems provide robust validation of the implementation.
In this section we first test the modules and the code in the Newtonian regime, then present general-relativistic examples in Section~\ref{sec:application_tests}.

\subsection{\label{sec:eps_test}Specific energy to temperature inversion}
The specific energy $\epsilon$ is required for hydrodynamical evolutions, while the EoSs typically use temperature $T$ as an input rather than $\epsilon$.
Calculating the specific energy from a given temperature (i.e., $\epsilon(T)$) is straightforward.
However, inverting from specific energy to temperature (i.e., $T(\epsilon)$) is non-trivial and often requires a root-finding procedure.

Here, we test our implementation of the specific energy-to-temperature inverter.
The workflow of the round-trip test is
\begin{equation}
	T \xrightarrow{\text{EoS}} \epsilon
	 \xrightarrow{\text{$\epsilon$-$T$ root finding}} T' \xrightarrow{\text{EoS}} \epsilon'.
\end{equation}
The relative error of $T$ is then calculated as
\begin{equation}
	\text{relative error of } T = \left\lvert \frac{T'}{T}-1 \right\rvert,
\end{equation}
with a similar definition for the relative error of $\epsilon$.

We sample the rest-mass density and temperature over $-10 \leq \log_{10}\left(\rho~{\rm [g \cdot cm^{-3}]}\right) \leq 15$ and $4 \leq \log_{10}\left(T~{\rm [K]}\right) \leq 12$, using 32 points in each direction.
The electron fraction $Y_{\rm e}$ is set to $\{0.5, 0.4, 0.3, 0.2, 0.1\}$.
The SFHo EoS~\citep{2013ApJ...774...17S} is used for the NSE region.
For non-NSE regions, mass fractions $\{X_l\}$ are required.
Instead of assuming NSE everywhere, we assign $\{X_l\}$ based on a ``two-species mixture'' for a given $Y_{\rm e}$.
Specifically, we scan from the lightest to heaviest isotopes and identify two isotopes $i$ and $j$ such that $Z_i/A_i \leq Y_{\rm e} \leq Z_j/A_j$.
The mass fractions are then assigned as
\begin{equation}
	X_i = \frac{Y_{\rm e}-Z_j/A_j}{Z_i/A_i-Z_j/A_j}, \qquad X_j = 1 - X_i.
\end{equation}

\begin{figure*}
	\centering
	\includegraphics[width=\textwidth, angle=0]{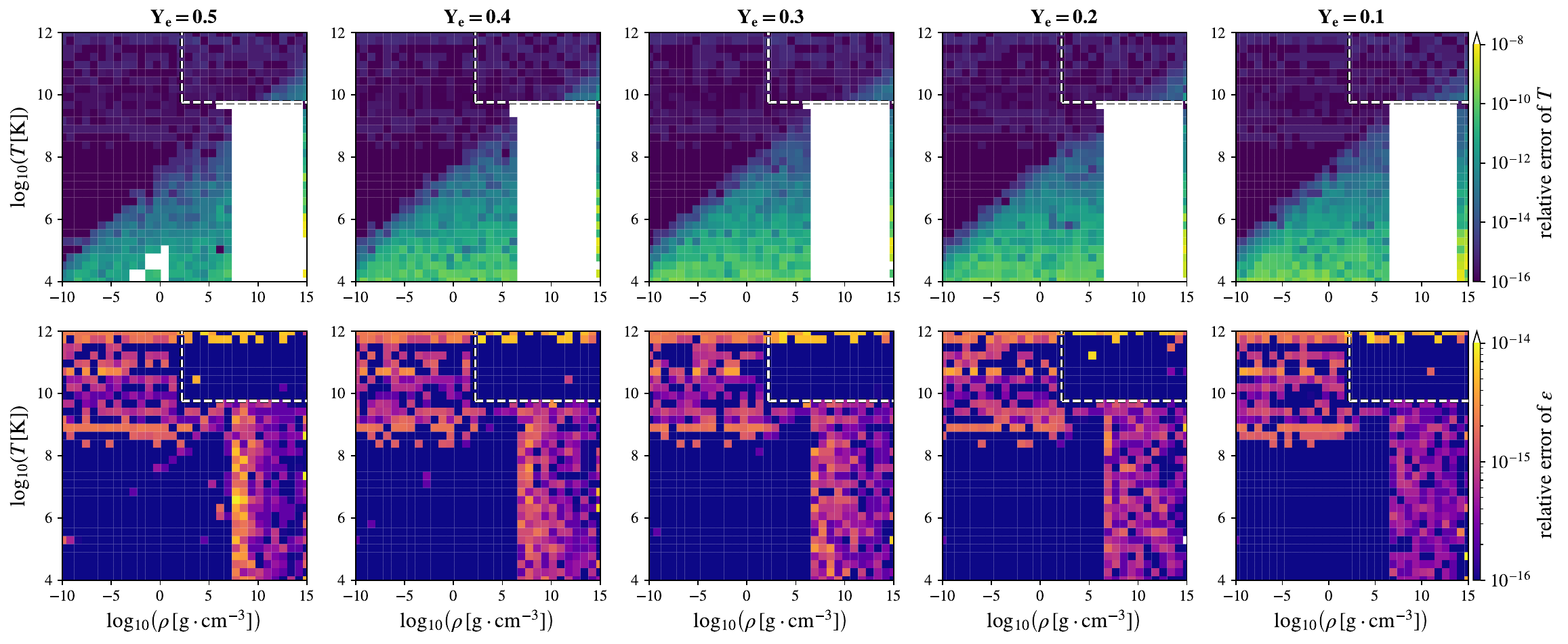}
	\caption{
		Relative errors of temperature $T$ (\emph{top}) and specific energy $\epsilon$ (\emph{bottom}) from the round-trip tests for five electron fractions $Y_{\rm e}$.
		The upper-right regions enclosed by the white dashed lines denote where the nuclear EoS is applied, while the stellar or transitional (mixed) EoS is used elsewhere.
		The mixed regime corresponds to $5 < T < 5.8~{\rm GK}$ where both the stellar and nuclear EoS tables overlap in density.
		At densities $\rho \gtrsim 10^{7}~{\rm g~cm^{-3}}$ and temperatures $T \lesssim 10^{10}~{\rm K}$, the temperature recovery is imperfect because the code prioritizes solutions from the nuclear EoS (see Section~\ref{sec:eos_bridging}).
		Nevertheless, the relative error in the specific energy $\epsilon$, which serves as a primitive variable in hydrodynamical evolution, remains below $10^{-14}$ across the explored parameter space.
	}
	\label{fig:c2p_T_eps_err}
\end{figure*}

Figure~\ref{fig:c2p_T_eps_err} shows the relative errors of temperature $T$ and specific energy $\epsilon$ for the round-trip tests with different electron fractions $Y_{\rm e}$.
As discussed in Section~\ref{sec:eos_bridging}, the solution $T$ of $\epsilon = \epsilon(T)$ may exist in both EoSs in regions where $\rho \gtrsim 10^{7}~{\rm g \cdot cm^{-3}}$ and $T \lesssim 10^{10}~{\rm K}$.
Since the code prioritizes the solution available in the nuclear EoS, the temperature is not exactly recovered in these regions, which corresponds to the white areas in the lower-right part of the figures.
Nevertheless, the relative error of the specific energy in these regions remains below $10^{-14}$ (bottom panels).
Therefore, as long as no microphysics explicitly requires the temperature in these regions, the hydrodynamical evolution is expected to remain accurate.

\subsection{\label{sec:c2p_test}Conserved-to-primitive conversion}
Recovery of primitive variables from conservative variables in GRMHD is a non-trivial process, requiring the numerical solution of non-linear equations.
As discussed in Section~\ref{sec:eos}, \texttt{Gmunu} adopts the conserved-to-primitive conversion scheme of \cite{2021PhRvD.103b3018K}.
Although this scheme has been successfully applied with tabulated nuclear EoSs, it warrants re-testing due to the special bridging between stellar and nuclear EoSs.

The primitive vector is defined as
\begin{equation}\label{eq:prim}
	{\bm p} = \left[ \rho, Wv^i, \log\left(T\right), B^i, Y_{\rm e}, X_{l} \right].
\end{equation}
To assess the solver performance, we sample the primitive variables over a specified range, compute the corresponding conservative variables, then recover the primitive variables and compare them to the original vector.
The accuracy is quantified via \citep{2018ApJ...859...71S}
\begin{equation}
	{\rm error} = \frac{1}{N_{\rm p}} \sum_{i=1}^{N_{\rm p}} \left\lvert \frac{ {\bm p}_i - {\bm p}'_{i}}{{\bm p}'_{i}} \right\rvert,
\end{equation}
where $N_{\rm p}$ is the number of primitive variables.

While the primitive vector often uses the specific energy $\epsilon$ rather than temperature $T$ (e.g., \cite{2018ApJ...859...71S}), we measure the error using $T$ because it is required for many microphysics calculations, such as nuclear burning and neutrino transport.
Consequently, relative errors with $T$ are generally larger than those with $\epsilon$ (see Figure~\ref{fig:c2p_T_eps_err}).

\begin{figure*}
	\centering
	\includegraphics[width=\textwidth, angle=0]{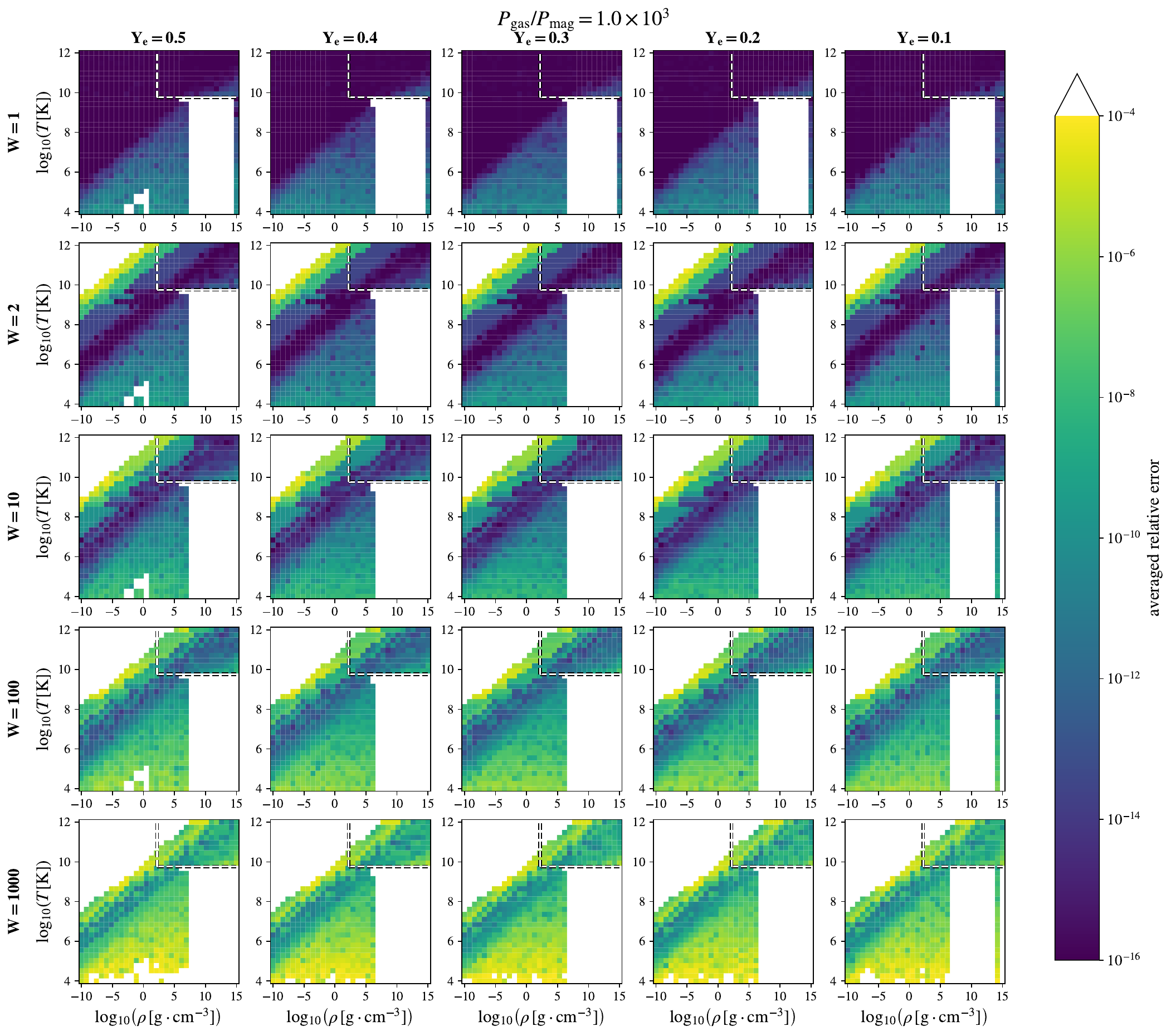}
	\caption{
		Averaged relative error in conserved-to-primitive tests for selected Lorentz factors $W$, electron fractions $Y_{\rm e}$, and magnetization $\beta_{\rm mag} = 1000$ (weakly magnetized case).
		The upper-right regions enclosed by the white dashed lines denote where the nuclear EoS is applied, while the stellar or transitional (mixed) EoS is used elsewhere.
	}
	\label{fig:c2p_beta1e3}
\end{figure*}

Figure~\ref{fig:c2p_beta1e3} shows the averaged error in conserved-to-primitive conversions, sampling $\rho$ and $T$ for selected Lorentz factors $W$, electron fractions $Y_{\rm e}$, and magnetization $\beta_{\rm mag} = P_{\rm gas}/P_{\rm mag} = 1000$ (weakly magnetized case).
White regions in the lower right reflect temperature recovery issues discussed in Section~\ref{sec:eps_test}.
The upper-left region (low density, high temperature) exhibits larger errors for higher Lorentz factors because, in this regime, kinetic energy dominates the conserved variables, making the solution insensitive to thermal energy and causing precision loss due to floating-point cancellation.
Tests with stronger magnetic fields (down to $\beta_{\rm mag} = 10^{-3}$) show similar behavior and are not shown.
Overall, the majority of the parameter space exhibits relative errors below $10^{-8}$, with errors near the edges reaching $\lesssim 10^{-4}$.

\subsection{Pure Silicon Burning Test}
To verify that the nuclear burning module drives compositions consistently toward NSE, we performed a one-zone calculation starting from pure $\rm ^{28}Si$. 
This setup has long been used as a benchmark in nuclear network studies (e.g., \cite{1973ApJS...26..231W, 1996ApJ...460..869H, 1999JCoAM.109..321H, 2000ApJS..129..377T}), since silicon burning rapidly produces an iron-group dominated composition under high temperatures and densities.

\begin{figure}
	\centering
	\includegraphics[width=\columnwidth]{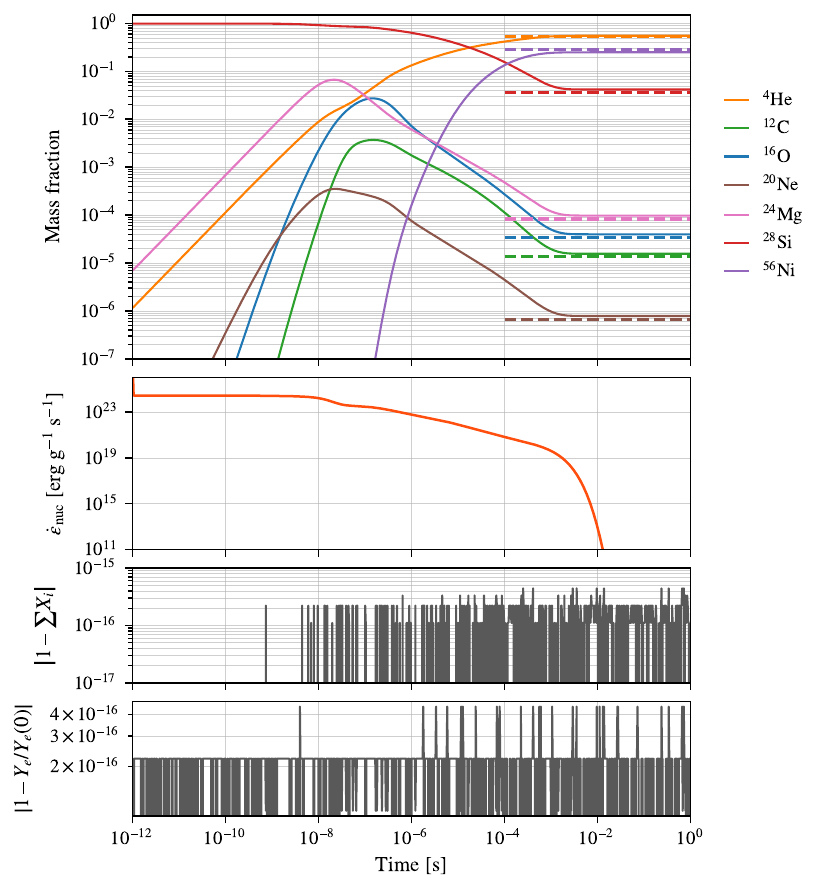}
	\caption{
		One-zone pure silicon burning test at fixed $\rho = 10^{7}~\rm{g \cdot cm^{-3}}$ and $T = 6\times10^{9}~{\rm K}$, starting from pure $\rm ^{28}Si$. 
		The network evolves the composition toward an iron-group dominated state consistent with nuclear statistical equilibrium. 
		\emph{Top panel:} Mass fractions of selected species (solid) and NSE predictions (dashed). 
		\emph{Middle panel:} Energy generation rate. 
		\emph{Bottom panels:} Deviations from conservation: $\left\vert 1 - \sum X_l \right\rvert$ and $\left\vert 1 - Y_{\rm e}/Y_{\rm e}(0) \right\rvert$.
	}
	\label{fig:si_burning_test}
\end{figure}

We fix the thermodynamic conditions at $\rho = 10^{7}~{\rm g \cdot cm^{-3}}$ and $T = 6~{\rm GK}$, and evolve the composition using our nuclear reaction network.
The evolution is performed over a time domain sampled uniformly in $\log_{10}$ space from $10^{-12}$ to $1~{\rm s}$ with 1000 points, which directly sets the integration timesteps.  
Thanks to the robustness of the Netwon-Raphson solver, the test proceeds smoothly even under this aggressive timestep choice.  
Weak interactions are neglected, and therefore the electron fraction $Y_{\rm e}$ remains constant throughout the calculation.  
The system evolves until the net reaction flows approach detailed balance, indicating relaxation to nuclear statistical equilibrium (Figure~\ref{fig:si_burning_test}).

To ensure the numerical integrity of the network, we monitor conservation properties:
\begin{itemize}
    \item \textbf{Mass fraction sum:} $\sum_l X_l$ should remain unity, as the network only redistributes composition without creating or destroying total nucleon number.
    \item \textbf{Electron fraction:} $Y_{\rm e}$ must remain constant in the absence of weak interactions, providing an independent check of numerical consistency.
\end{itemize}

The final abundances are compared with predictions from the NSE solver at the same thermodynamic state. 
The network converges toward the expected equilibrium distribution. 
The integrated energy release during burning matches the binding energy differences implied by the evolving composition, confirming the correct coupling of the network to the equation of state.

The NSE abundances and the asymptotic abundances from the network calculation show slight differences.
This arises because the Coulomb physics treated in the NSE solver and in the {\tt aprox13} network are not identical.
The NSE module incorporates Coulomb corrections following \cite{1998PhRvE..58.4941C}.
In contrast, {\tt aprox13} accounts for Coulomb effects through screening corrections to reaction rates based on \cite{1982ApJ...258..696W}.

This test demonstrates that the network reliably drives material from a non-equilibrium initial composition to NSE under fixed conditions, establishing a controlled baseline for subsequent tests that include hydrodynamic and relativistic coupling. 
In particular, conservation of $\sum X_l$ and $Y_{\rm e}$ in this simple setup builds confidence that these quantities will remain robust when the network is coupled to full multidimensional simulations, where violations could otherwise accumulate and compromise the long-term stability of the calculation.

\subsection{Newtonian Real Gas Shock Tubes}
To verify the stellar EoS implementation, we perform the Newtonian real gas shock tube test introduced by \cite{2015ApJS..216...31Z}, for which exact solutions are available. 
The simulations are carried out in Cartesian coordinates.

Following test~1 of \cite{2015ApJS..216...31Z}, the initial conditions are
\begin{align}
	\left( \rho, T, v \right) = 
	\begin{cases}
		\left( \rho_{\rm L}, T_{\rm L}, v_{\rm L}\right),	&	\text{if } x < 0.5 L, \\
		\left( \rho_{\rm R}, T_{\rm R}, v_{\rm R}\right),	&	\text{if } x > 0.5 L,
	\end{cases}
\end{align}
with parameters
\begin{align}
	& \rho_{\rm L} = 10^{7}~{\rm g \cdot cm^{-3}}, && \rho_{\rm R} = 10^{6}~{\rm g \cdot cm^{-3}}, \\
	& T_{\rm L} = 10^{8}~{\rm K}, && T_{\rm R} = 10^{6}~{\rm K}, \\
	& v_{\rm L} = 0, && v_{\rm R} = 0,
\end{align}
and domain length $L = 10^{6}~\rm{cm}$. 
The composition is set to pure $\rm ^{12}C$.

\begin{figure*}
	\centering
	\includegraphics[width=\textwidth]{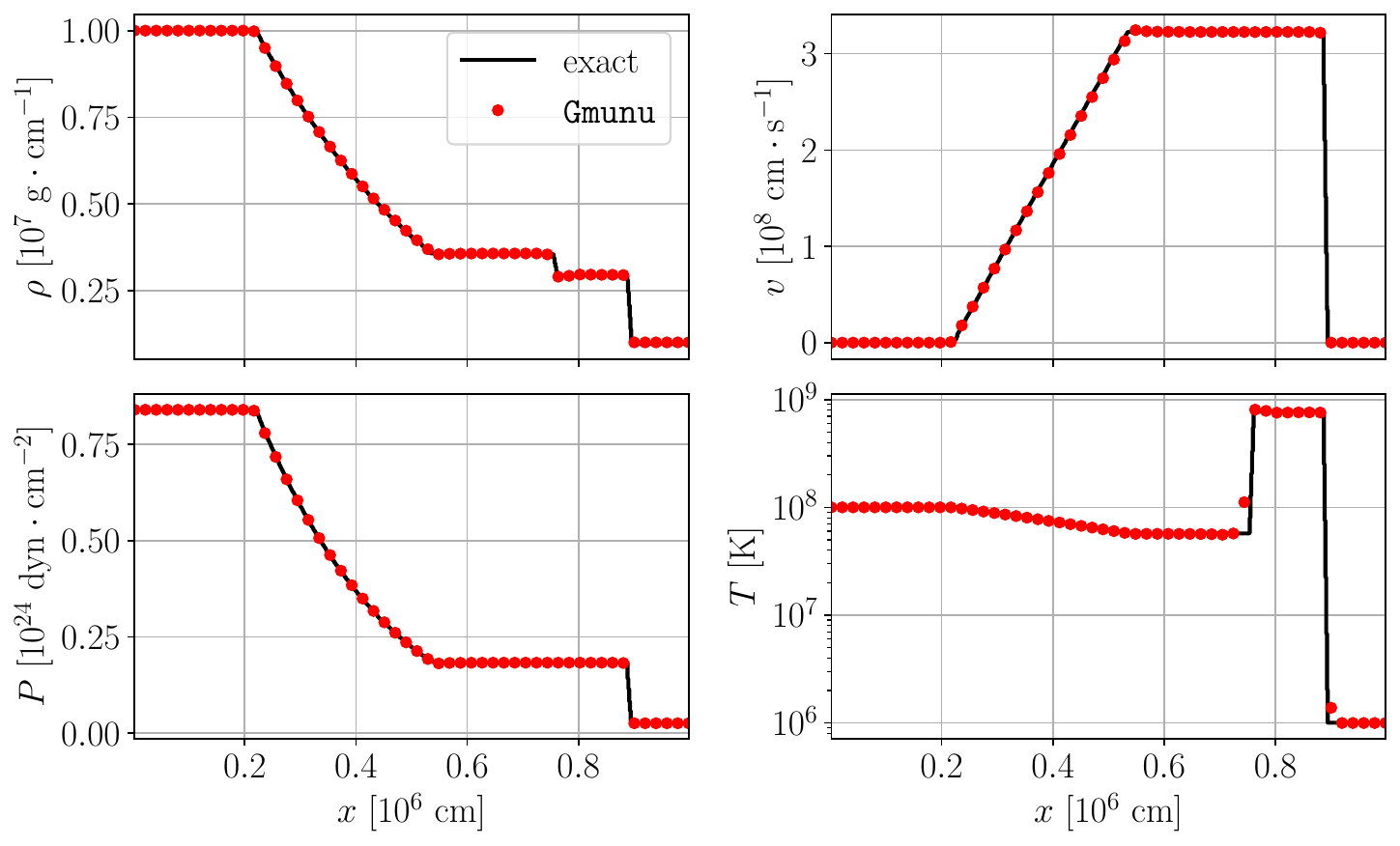}
	\caption{
		Profiles of rest-mass density $\rho$ (\emph{upper left}), velocity $v$ (\emph{upper right}), pressure $P$ (\emph{lower left}), and temperature $T$ (\emph{lower right}) at $t = 8\times10^{-4}~{\rm s}$ for the Newtonian real gas shock tube test~1 of \cite{2015ApJS..216...31Z}. 
		Red dots: numerical results from \texttt{Gmunu}; black solid lines: exact reference solution. 
		The excellent agreement validates the stellar EoS implementation.
	}
	\label{fig:realgas_shocktube_1d}	
\end{figure*}

Figure~\ref{fig:realgas_shocktube_1d} shows the profiles of rest-mass density $\rho$, velocity $v$, pressure $P$, and temperature $T$ at $t = 8\times10^{-4}~{\rm s}$. 
The numerical results from \texttt{Gmunu} (red dots) agree well with the reference exact solutions (black solid lines), confirming the correct implementation of the stellar EoS in Newtonian regimes.

\subsection{Newtonian Real Gas Acoustic Pulse in 1D}

Smooth tests are ideal for assessing the convergence of the implementation. 
Here, we perform the Newtonian real gas acoustic pulse test introduced in \cite{2019ApJ...886..105Z} to validate both hydrodynamics and nuclear burning in \texttt{Gmunu}. 
The initial pressure and specific entropy profiles are prescribed as:
\begin{align}
	p\left(x\right) &= p_0 \left[ 1 + f_p \exp\left(-\frac{x^2}{\delta_r^2}\right) \cos^6\left(\frac{\pi x}{L_x}\right) \right], \\
	s\left(x\right) &= s_0,
\end{align}
where $f_p$ and $\delta_r$ set the amplitude and width of the perturbation, $L_x$ is the domain size in the $x$-direction, and $x$ is the spatial coordinate. 
The ambient pressure $p_0$ and specific entropy $s_0$ are determined from the background density $\rho_0$ and temperature $T_0$, i.e., $p_0 = p(\rho_0, T_0)$ and $s_0 = s(\rho_0, T_0)$.  

Following \cite{2025ApJ...981...63H}, we adopt $\rho_0 = 5 \times 10^{5}~{\rm g \cdot cm^{-3}}$, $T_0 = 3 \times 10^8~{\rm K}$, $f_p = 2$, $\delta_r = 2 \times 10^7~{\rm cm}$, and $L_x = 10^8~{\rm cm}$.  
The fluid is initially composed of pure $\rm ^4He$.  

We employ the 2nd-order accurate strong-stability preserving IMEX-SSP2(2,2,2) time integrator \citep{pareschi2005implicit}, the Harten-Lax-van Leer (HLL) Riemann solver, and a 2nd-order monotonized central (MC) limiter \citep{1974JCoPh..14..361V}. 
For the nuclear-reacting cases, the 13-species network {\tt aprox13} is coupled to the hydrodynamics.

\begin{figure}
	\centering
	\includegraphics[width=\columnwidth]{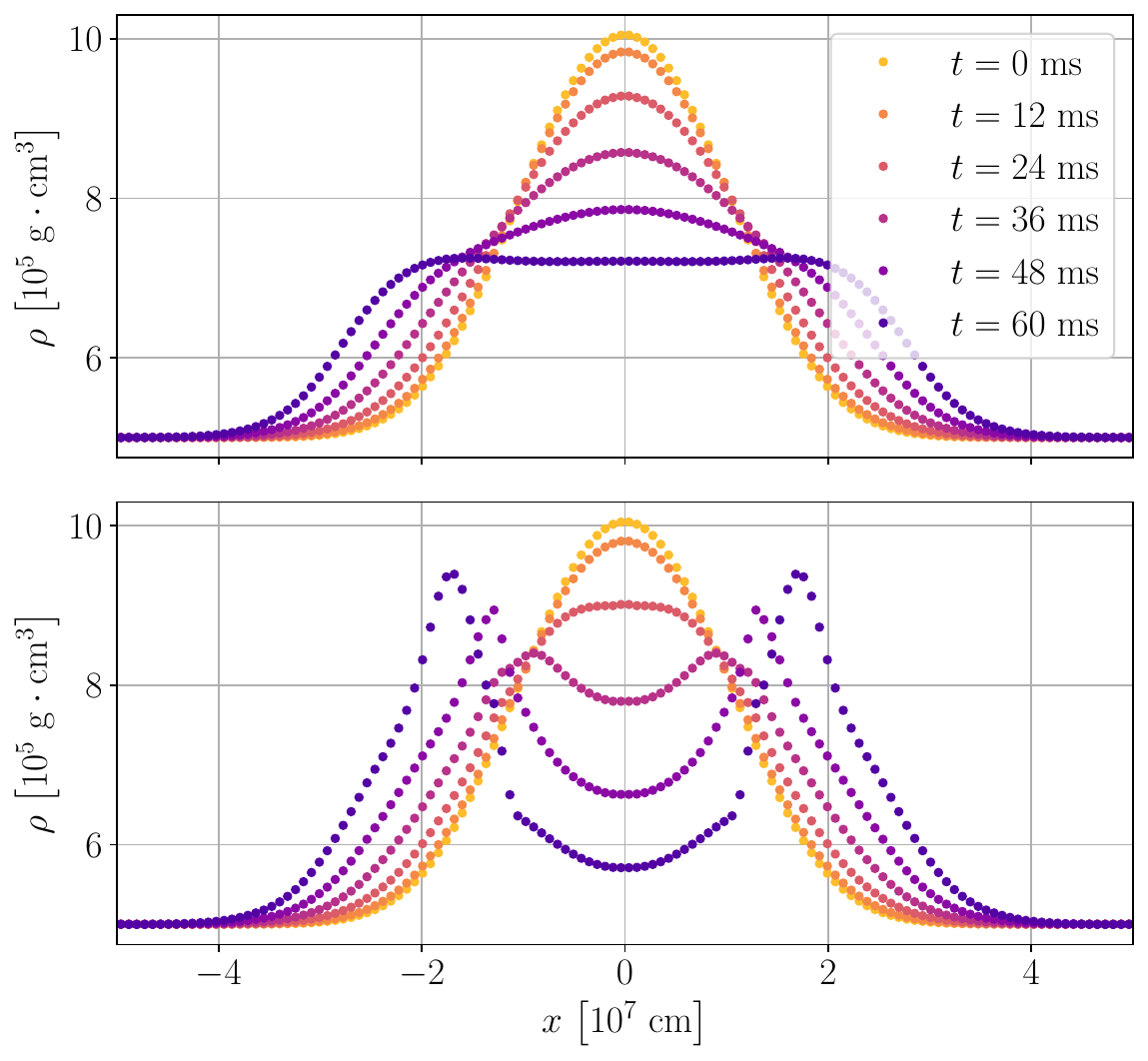}
	\caption{
		Snapshots of the rest-mass density $\rho$ at six times for the Newtonian real gas acoustic pulse test \citep{2019ApJ...886..105Z, 2025ApJ...981...63H}. 
		The upper panel shows pure hydrodynamics, and the lower panel includes nuclear reactions via the 13-species network {\tt aprox13}. 
		When nuclear burning is included, the expansion dynamics are visibly altered.
	}
	\label{fig:acoustic_pulse_1d}
\end{figure}

Figure~\ref{fig:acoustic_pulse_1d} shows snapshots of the rest-mass density $\rho$ at six times for $N=128$ grid points. 
The comparison between purely hydrodynamical evolution (upper panel) and the nuclear-reacting case (lower panel) demonstrates that nuclear burning significantly affects the expansion dynamics of the fluid.

To quantify convergence, we perform simulations at multiple spatial resolutions and compute the volume-weighted $L_1$ norm of the rest-mass density relative to a high-resolution reference solution ($N_{\rm ref} = 2048$):
\begin{equation}
	L_1\left(N\right) = 
	\frac{\sum_{i=1}^{N} \Delta x[i] \times \left \vert \rho[i] - \rho_{\rm ref}\left[ \frac{N_{\rm ref}}{N} \times i \right] \right\vert}
	     {\sum_{i=1}^{N} \Delta x[i]},
\end{equation}
where $\Delta x[i]$ is the local cell width and $i$ indexes the grid points. 
The diagnostic is evaluated at $t_{\rm f} = 0.02~{\rm s}$. 
A fixed timestep scaled with resolution is adopted following \citet{2019ApJ...886..105Z}:
\begin{equation}
	\Delta t = 2 \times 10^{-4} \left( \frac{64}{N} \right)~{\rm s}.
\end{equation}
As discussed by \citet{2025ApJ...981...63H}, entropy is not conserved when nuclear burning is active; therefore, we monitor convergence using the rest-mass density instead.

\begin{figure}
	\includegraphics[width=\columnwidth]{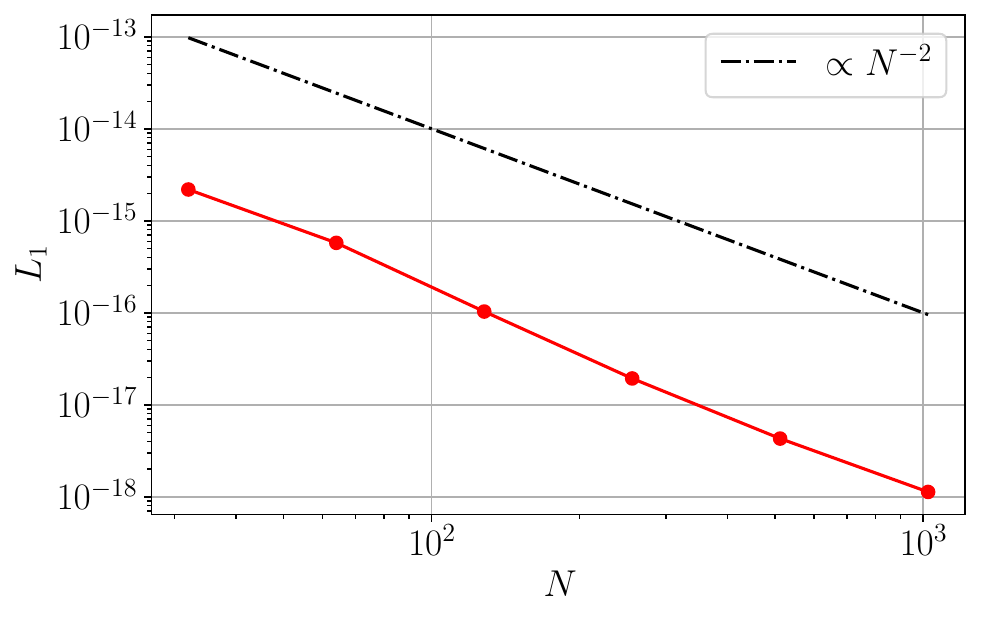}
	\caption{
		Volume-weighted $L_1$-norm versus resolution $N$ for the Newtonian real gas acoustic pulse at $t = 0.02~{\rm s}$.  
		Second-order ideal scaling is shown with a dashed line, confirming that the implementation achieves the expected convergence rate.
	}
	\label{fig:acoustic_pulse_1d_convergence_rate}
\end{figure}

Figure~\ref{fig:acoustic_pulse_1d_convergence_rate} shows $L_1$ as a function of resolution $N$.  
Second-order convergence is observed, in agreement with the 2nd-order time integrator and spatial reconstruction.

\subsection{\label{sec:burning_shock_test}Detonation front of SNe Ia in 1D}
To validate our burning limiter prescription, we reproduce a test problem originally proposed by \cite{2014ApJ...782...12P}, which examines detonations in the outer reactive layers of Type~Ia supernovae.  
This test provides a clean setup for studying the coupling between hydrodynamic shocks and nuclear burning.

The 2nd-order accurate strong-stability preserving IMEX-SSP2(2,2,2) time integrator \citep{pareschi2005implicit}, the Harten-Lax-van Leer (HLL) Riemann solver, and a 2nd-order monotonized central (MC) limiter \citep{1974JCoPh..14..361V}, are employed in this test. 
The computational domain spans $\left[0, 4096\right]~{\rm cm}$ and is discretized with 256 base zones, allowing up to six levels of adaptive mesh refinement (AMR) that track the burning fronts of the ${}^{12}{\rm C}$ and ${}^{16}{\rm O}$ mass fractions.  
At the highest refinement level, the effective grid spacing is $\Delta x = 0.5~{\rm cm}$.  
Reflective boundary conditions are applied at the left boundary (ash side), while outflow boundary conditions are imposed at the right boundary (fuel side).  
The unperturbed region initially has a uniform rest-mass density and temperature of $\rho = \rho_0$ and $T = 10^{7}~{\rm K}$, respectively, with equal mass fractions of ${}^{12}{\rm C}$ and ${}^{16}{\rm O}$ only.
We consider four density setups with $\rho_0 = 5\times10^{7}$, $1\times10^{8}$, $5\times10^{8}$, and $1\times10^{9}~{\rm g \cdot cm^{-3}}$.  
To initiate the detonation, we set $T = 10^{10}~{\rm K}$ and $v^x = 10^{9}~{\rm cm \cdot s^{-1}}$ for $x < 25~{\rm cm}$.

We perform simulations both with and without the burning limiter mechanism described in Section~\ref{sec:coupling_to_hydro}.  
In the runs with suppression enabled, nuclear burning is deactivated in zones flagged as shocked, thereby preventing premature or spurious nuclear energy release at the shock front.

\begin{figure}
	\centering
	\includegraphics[width=\columnwidth, angle=0]{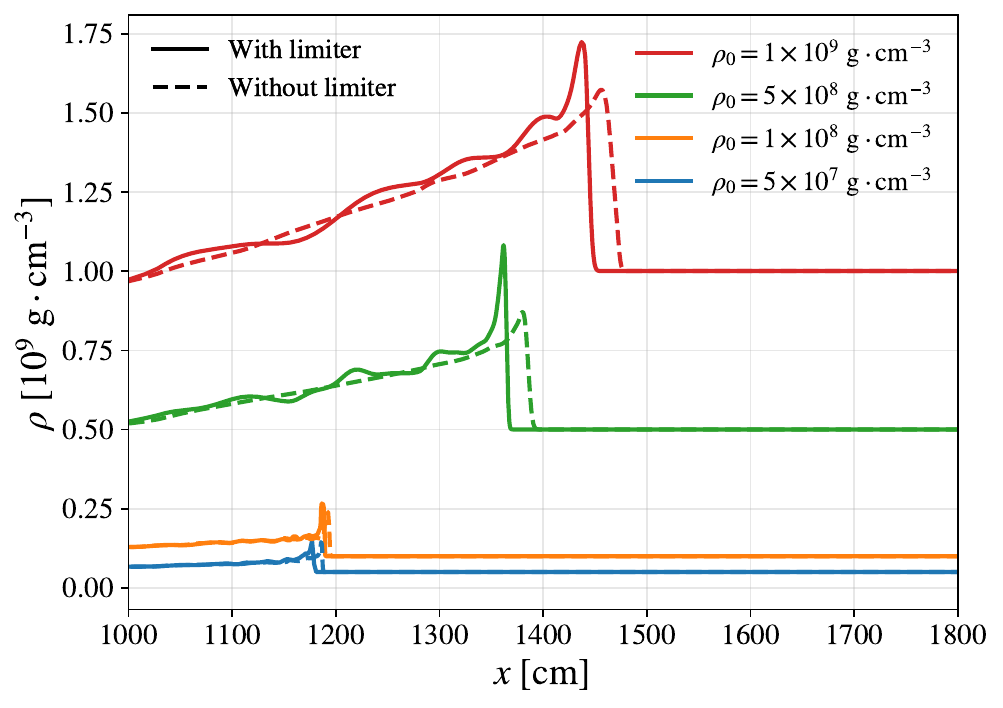}
	\caption{
		Rest-mass density profiles $\rho$ at $t = 10^{-6}~{\rm s}$ for different models with (solid lines) and without (dashed lines) burning suppression at shocks.
		We qualitatively reproduce the results of \cite{2014ApJ...782...12P}, showing that shocks propagate faster when burning is not suppressed.
		Minor offsets in shock position are expected due to slight differences in the evaluation time and numerical treatment. 
		Unlike in \cite{2014ApJ...782...12P}, we do not observe the secondary shock feature, which may be related to the IMEX integration scheme used here.
		}
	\label{fig:burning_shock_rho_profiles}
\end{figure}

Figure~\ref{fig:burning_shock_rho_profiles} shows the rest-mass density profiles at $t = 10^{-6}~{\rm s}$ for models with and without burning suppression.  
Solid lines denote cases where burning is prohibited in shocked regions, while dashed lines correspond to runs where burning is always allowed.  
Overall, our results qualitatively reproduce the trends reported by \cite{2014ApJ...782...12P}: cases without suppression develop slightly faster-moving shocks due to additional energy deposition from unphysical burning immediately behind the shock.  
The absolute shock locations differ marginally from those in their study, likely due to differences in the precise output time (their figure does not show results exactly at $t = 10^{-6}~{\rm s}$) and in the numerical schemes used.

However, we do not observe the secondary shock structure reported by \cite{2014ApJ...782...12P}.  
This discrepancy may stem from our use of an IMEX time-integration scheme with direct Newton–Raphson coupling between hydrodynamics and nuclear burning, whereas most previous nuclear hydrodynamics codes evolve the reaction network via operator-split ODE solvers.  
In addition, differences in numerical dissipation—arising from less diffusive Riemann solvers, reconstruction methods, or the AMR configuration—may also contribute.  
The piecewise-parabolic method \citep{1984JCoPh..54..174C} used by \cite{2014ApJ...782...12P} is generally less diffusive and more accurate in smooth regions than the MC reconstruction adopted here.
However, because the time-integration scheme and Riemann solvers employed in their work are not documented, a more detailed comparison with \cite{2014ApJ...782...12P} is not possible here.
A more detailed investigation is warranted to assess how different coupling strategies (fully coupled vs.\ operator-split) influence post-shock burning dynamics and the formation of secondary shocks.

\section{\label{sec:application_tests}Application example: CCSNe in 1D}
Core-collapse supernovae (CCSNe) provide highly demanding test beds due to their multi-physics and multi-scale nature.
In this section, we present spherically symmetric (1D) simulations of CCSNe with nuclear burning.

We consider the collapse of two solar-metallicity progenitors: a 9~M$_\odot$ star \citep{2016ApJ...821...38S} and a 20~M$_\odot$ star \citep{2007PhR...442..269W}.
The SFHo EoS~\citep{2013ApJ...774...17S} is adopted for the NSE regime.
Time integration is performed with the IMEX-SSP2(2,2,2) scheme \citep{pareschi2005implicit}, and spatial reconstruction uses the second-order monotonized central (MC) limiter \citep{1974JCoPh..14..361V}. 
The computational domain extends over $r \in [0,10^5]~{\rm km}$ with $N_r = 128$ and up to $l_{\max}=13$ mesh-refinement levels, yielding a finest grid spacing of $\Delta r \approx 191~{\rm m}$.
Refinement criteria follow \cite{2023ApJS..267...38C}.

Neutrino transport is treated with the energy-dependent two-moment scheme of \cite{2023ApJS..267...38C}. 
Microphysics is provided by \texttt{NuLib}~\citep{2015ApJS..219...24O}, with interaction rates consistent with \cite{2018JPhG...45j4001O}. 
The energy space is discretized into 18 logarithmic groups spanning $1$--$280~{\rm MeV}$. 
Since \texttt{NuLib} requires a nuclear EoS under the assumption of NSE, neutrino interactions are included only in regions with $T > 5~{\rm GK}$. 
While this introduces a simplification at lower temperatures, it remains a reasonable choice for the CCSN dynamics studied here. 
Nevertheless, a more general treatment that accounts for non-NSE effects will be pursued in future work.

Free neutrons and protons are included as advected species but do not participate in the 13-isotope $\alpha$-chain reactions.
Their inclusion ensures that the composition can represent an arbitrary $Y_{\rm e}$ when material cooled by neutrino processes enters the low-temperature regime where the network is active.
It also allows the NSE solver to operate consistently within the mixed EoS region ($5 < T < 5.8~{\rm GK}$).
A more complete network that explicitly includes neutrons, protons, and heavier nuclei such as ${}^{56}\mathrm{Fe}$ would remove this limitation; extending the reaction set in this direction is planned for future work.

In the post–bounce phase of a CCSN, the gain region is the layer between the stalled accretion shock and the neutrinosphere where net neutrino heating occurs.
In this zone, absorption of electron-type neutrinos dominates over neutrino cooling, resulting in a positive net heating rate.
Matter advected through the gain layer can acquire sufficient thermal and pressure support to aid shock revival, making it a key diagnostic for assessing the efficiency of the neutrino-driven explosion mechanism.
Following this physical picture, we estimate the net neutrino heating rate by
\begin{equation}\label{eq:nu_heat}
	\dot{Q}_\nu = \int_{\rm gain \, region} \left[ \int s_{\nu} \, \dd V_{\nu} \right] \sqrt{\gamma} \, \dd V,
\end{equation}
where $s_{\nu}$ denotes the neutrino–matter energy coupling source term~\citep{2023ApJS..267...38C}, $\dd V_{\nu}$ is the volume element in energy space, and $\dd V$ the spatial volume element. 
The gain region is defined as the region where $\rho < 3 \times 10^{10}~\mathrm{g \cdot cm^{-3}}$, the entropy per baryon exceeds $6\,k_\mathrm{B}$, and $\int s_{\nu} \, \dd V_{\nu} > 0$.

The diagnostic explosion energy is given by
\begin{equation}
	E_\mathrm{exp} = \int_{\varepsilon_\mathrm{exp}>0, \, v_r>0} \varepsilon_\mathrm{exp} \, \sqrt{\gamma} \, \dd V,
\end{equation}
where the energy density $\varepsilon_\mathrm{exp}$ is defined as
\begin{equation}\label{eq:eps_exp}
	\varepsilon_\mathrm{exp} = W \rho \left( -h_\mathrm{th} u_t - 1 \right).
\end{equation}
Here, $u_t = W(- \alpha + \beta_i v^i )$, $v_r$ is the radial velocity, and 
\begin{equation}
	h_{\rm th} = 1 + \left(\epsilon - \epsilon_0\right) + \frac{P}{\rho}
\end{equation}
is the specific enthalpy including only the thermal internal energy. 
In this expression, $\epsilon$ is the specific internal energy, and $\epsilon_0$ is the corresponding zero-point obtained for the same rest-mass density and electron fraction but at zero temperature~\citep{2020PhRvD.102l3015B, 2025ApJ...978L..38C}. 
By construction, $\varepsilon_\mathrm{exp}>0$ corresponds to the Bernoulli criterion for unbound material.\footnote{%
Strictly speaking, the Bernoulli criterion is not the only possible definition of unbound matter, and can differ from more conservative approaches that account for long-term binding to the proto-neutron star. 
Nevertheless, it remains the standard diagnostic in CCSN studies and facilitates direct comparison across different works.}
In the Newtonian limit, equation~\eqref{eq:eps_exp} reduces to the commonly used diagnostic explosion energy in Newtonian CCSN simulations, where the gravitational binding energy is encoded in $u_t$.

The mass accretion rate at radius $r$ is computed as
\begin{equation}
	\dot{M}(r) = \int_{S(r)} \rho W v^r \sqrt{\gamma} \, \dd A,
\end{equation}
where the integral is performed over the spherical surface $S(r)$.
These diagnostics are widely adopted in the CCSN community to quantify the interplay of neutrino heating, explosion energetics, and mass inflow, and thus provide a direct basis for comparing with previous studies in both Newtonian and relativistic frameworks.

Since spherically symmetric iron-core collapse supernova models usually do not explode without additional treatment, we employ enhanced neutrino heating to trigger shock revival \citep{1993A&A...273..106B, 1995ApJS..101..181W, 2012ApJ...757...69U, 2015ApJ...806..275P, 2016ApJ...818..124E, 2020ApJ...890..127C, 2021ApJ...912...29B, 2023MNRAS.524.4109G}.
Following \cite{2023MNRAS.524.4109G}, we multiply the energy--momentum and $Y_{\rm e}$ source terms from electron-type neutrinos in the gain region by a constant factor $f_{\rm heat}$.
% The gain region is defined by $\rho < 3 \times 10^{10}~{\rm g \cdot cm^{-3}}$, $s > 6~k_{\rm B}/{\rm baryon}$, and net heating $\dd \tau_{\nu_{\rm e},\bar{\nu}_{\rm e}} > 0$.
While \cite{2023MNRAS.524.4109G} reported explosions of the 20~M$_\odot$ model for $f_{\rm heat} \leq 2.4$, here we adopt $f_{\rm heat} = 2.8$ for all cases, consistent with \cite{2023ApJ...951..112N}.
We emphasize that this scheme is artificial and requires calibration; self-consistent explosions require multidimensional simulations.
Here it is used solely to probe the role of nuclear burning behind the shock.

The initialization of composition depends on thermodynamic conditions.
In high-density regions where NSE is expected, abundances are computed with the NSE solver (Section~\ref{sec:NSE_solver}).
In non-NSE regions, the nuclei present in the progenitor profiles may not match those in the adopted network.
We therefore map progenitor abundances to the network species and rescale mass fractions to enforce $\sum X_l = 1$.
The network then relaxes dynamically over the first few timesteps.

We perform six simulations: for each progenitor, (i) a baseline case, (ii) a run with enhanced neutrino heating, and (iii) a run with both enhanced heating and nuclear burning.

\begin{figure*}
	\centering
	\includegraphics[width=\textwidth, angle=0]{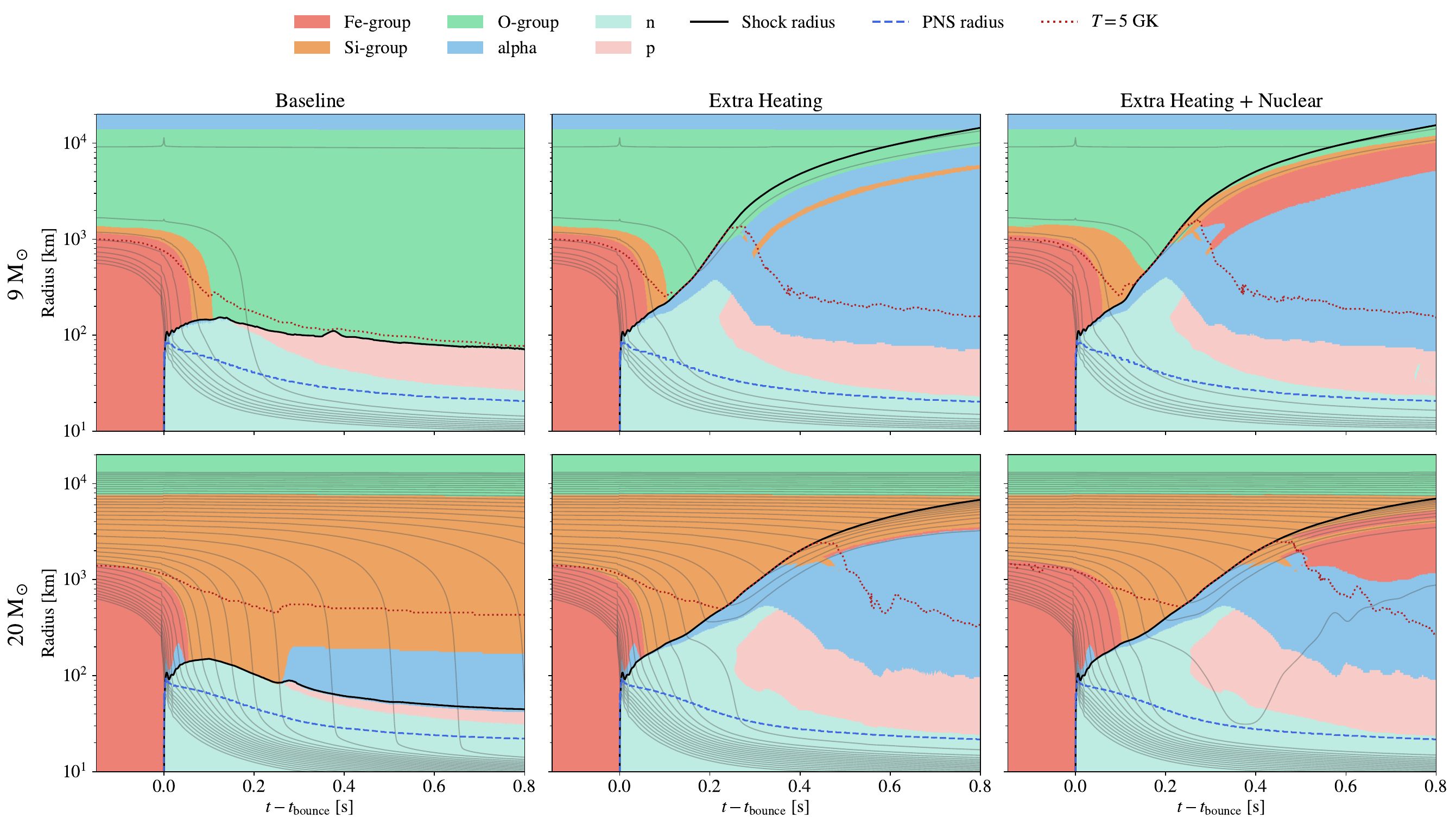}
	\caption{
		Temporal evolution of composition layers in the 9~M$_\odot$ (\emph{top}) and 20~M$_\odot$ (\emph{bottom}) progenitors under three setups (\emph{left to right}): baseline, enhanced neutrino heating, and heating plus nuclear burning. 
		Isotopes are grouped into six categories: 
		iron-group ($^{44}\mathrm{Ti},\,^{48}\mathrm{Cr},\,^{52}\mathrm{Fe},\,^{56}\mathrm{Ni}$), 
		silicon-group ($^{28}\mathrm{Si},\,^{32}\mathrm{S},\,^{36}\mathrm{Ar},\,^{40}\mathrm{Ca}$), 
		oxygen-group ($^{16}\mathrm{O},\,^{20}\mathrm{Ne},\,^{24}\mathrm{Mg}$), 
		alpha particle, neutron, and proton.
		Overplotted lines mark the shock radius, PNS surface, and temperature thresholds of $T=5~\mathrm{GK}$.
		Mass shells from 1 to 3~M$_\odot$ in increments of 0.05~M$_\odot$ are shown. 
		The figure highlights the role of neutrino heating and nuclear burning in modifying the post-shock composition during early post-bounce evolution.
		}
	\label{fig:nuc_ccsn_mass_shell_evolution}
\end{figure*}

Figure~\ref{fig:nuc_ccsn_mass_shell_evolution} summarizes the evolution of composition layers for the 9~M$_\odot$ and 20~M$_\odot$ progenitors under these setups.
For visualization, isotopes are grouped into six categories: 
iron-group ($^{44}\mathrm{Ti},\,^{48}\mathrm{Cr},\,^{52}\mathrm{Fe},\,^{56}\mathrm{Ni}$), 
silicon-group ($^{28}\mathrm{Si},\,^{32}\mathrm{S},\,^{36}\mathrm{Ar},\,^{40}\mathrm{Ca}$), 
oxygen-group ($^{16}\mathrm{O},\,^{20}\mathrm{Ne},\,^{24}\mathrm{Mg}$), 
alpha particle, neutron, and proton.
The dominant group (maximum mass fraction) is assigned in each grid cell.
Mass shells from 1 to 3~M$_\odot$ are overplotted in increments of 0.05~M$_\odot$ to illustrate advection and expansion.
Diagnostic quantities include the shock radius, proto-neutron star (PNS) surface, and temperature thresholds of $T=5~\mathrm{GK}$.
Note that \texttt{Gmunu} enforces NSE in the range $5$--$5.8~\mathrm{GK}$ (see Section~\ref{sec:NSE_solver}).

All baseline runs without enhanced heating or nuclear burning fail to explode.  
Our 20~M$_\odot$ baseline model is consistent with previous benchmarks obtained using different codes \citep{2018JPhG...45j4001O, 2024ApJS..272....9N, 2024ApJ...975..116C}.  
The 9~M$_\odot$ model has also been reported as non-exploding in one-dimensional simulations \citep{2017ApJ...850...43R}.  
Despite differences in microphysics and gravity treatments—\citet{2017ApJ...850...43R} adopted the stiffer LS220 EoS \citep{1991NuPhA.535..331L} and Newtonian gravity with an effective potential \citep{2006A&A...445..273M}, whereas our models employ full general relativity with the SFHo EoS—our results are qualitatively consistent.

All models with enhanced neutrino heating successfully explode.  
Although the shock and PNS radii evolve similarly across setups, the inclusion of nuclear burning significantly alters the post-shock composition.  
As shown in Figure~\ref{fig:nuc_ccsn_mass_shell_evolution}, when the post-shock matter cools below $5~{\rm GK}$, explosive Si/O burning produces additional iron-group nuclei.

\begin{figure*}
	\centering
	\includegraphics[width=\textwidth, angle=0]{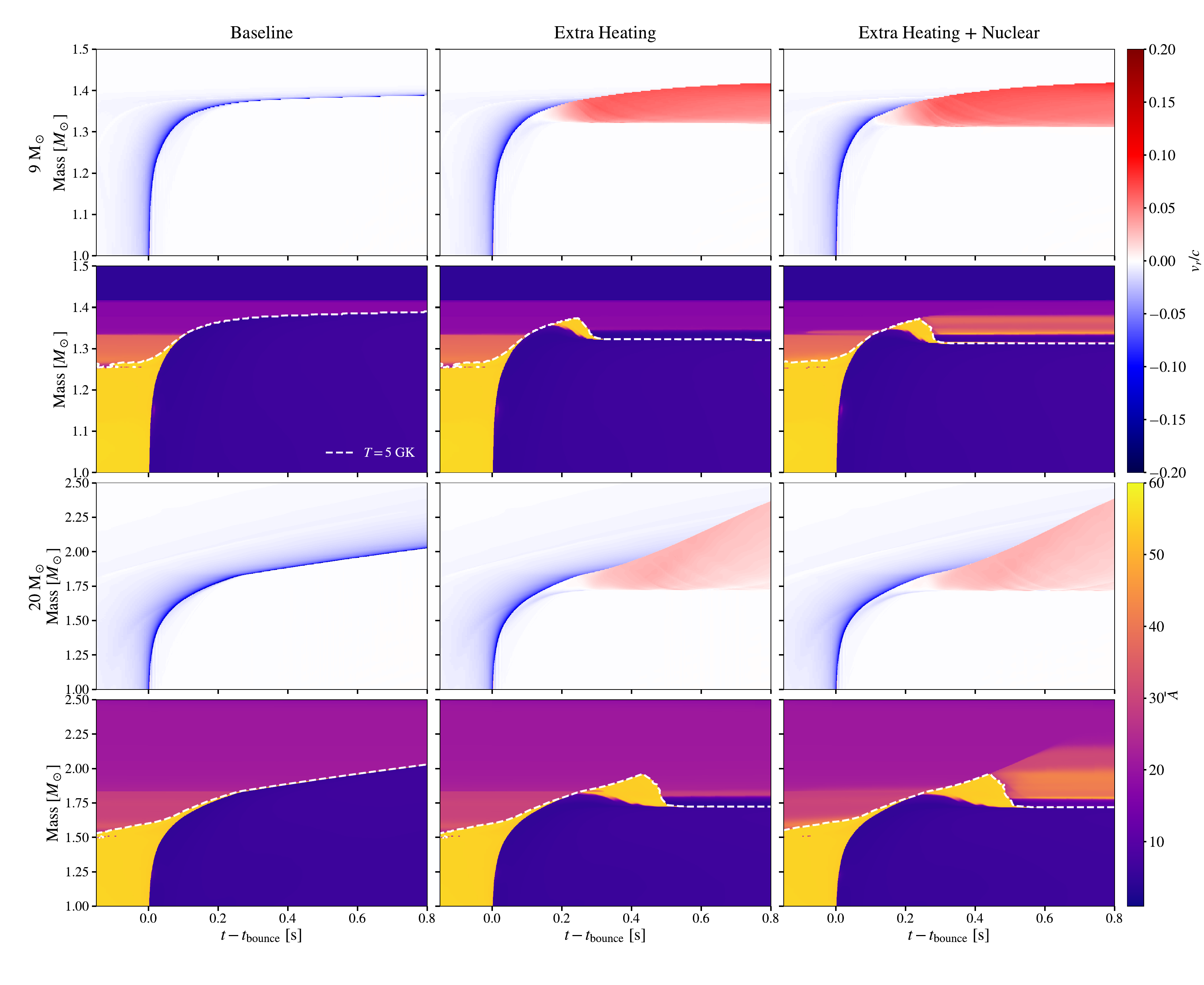}
	\caption{
		Time evolution of radial velocity $v_r$ (\emph{odd rows}) and mean mass number $\bar{A}$ (\emph{even rows}) in CCSN simulations of 9~M$_\odot$ (\emph{top}) and 20~M$_\odot$ (\emph{bottom}) progenitors. 
		Columns correspond to baseline (left), enhanced heating (middle), and heating plus burning (right).
		In baseline runs, the shock is not revived and composition remains largely unchanged.
		Enhanced heating prolongs shock expansion, but the post-shock matter is dominated by light nuclei. 
		With nuclear burning, $\bar{A}$ increases after a few 100~ms post-bounce as explosive burning drives material toward heavier nuclei. 
		The comparison shows that lower-mass progenitors undergo earlier composition changes, while higher-mass models require both heating and burning for significant enrichment.
		}
	\label{fig:nuc_ccsn_vr_abar_evolution}
\end{figure*}

Because assigning layers by dominant mass fraction is approximate, we also examine the radial velocity $v_r$ and mean mass number $\bar{A}$ in mass coordinates (Figure~\ref{fig:nuc_ccsn_vr_abar_evolution}).
For $T>5~{\rm GK}$, $\bar{A}$ is interpolated from the SFHo EoS; otherwise, it is computed directly from mass fractions.

At early times, high post-shock temperatures dissociate heavy nuclei (Si, Fe group) into free nucleons and $\alpha$ particles, reducing $\bar{A}$.
This dissociation absorbs $\sim 8~{\rm MeV}$ per nucleon, draining energy from the shock and causing it to stall.
As the shock weakens, material behind it cools faster than newly shocked matter heats, and the shock temperature eventually drops below $5~{\rm GK}$.
Without nuclear burning, compositions effectively freeze at this stage.
With burning included, regions with $T < 5~{\rm GK}$ undergo Si burning, steadily driving matter toward the iron group and increasing $\bar{A}$.

\begin{figure*}
	\centering
	\includegraphics[width=\textwidth, angle=0]{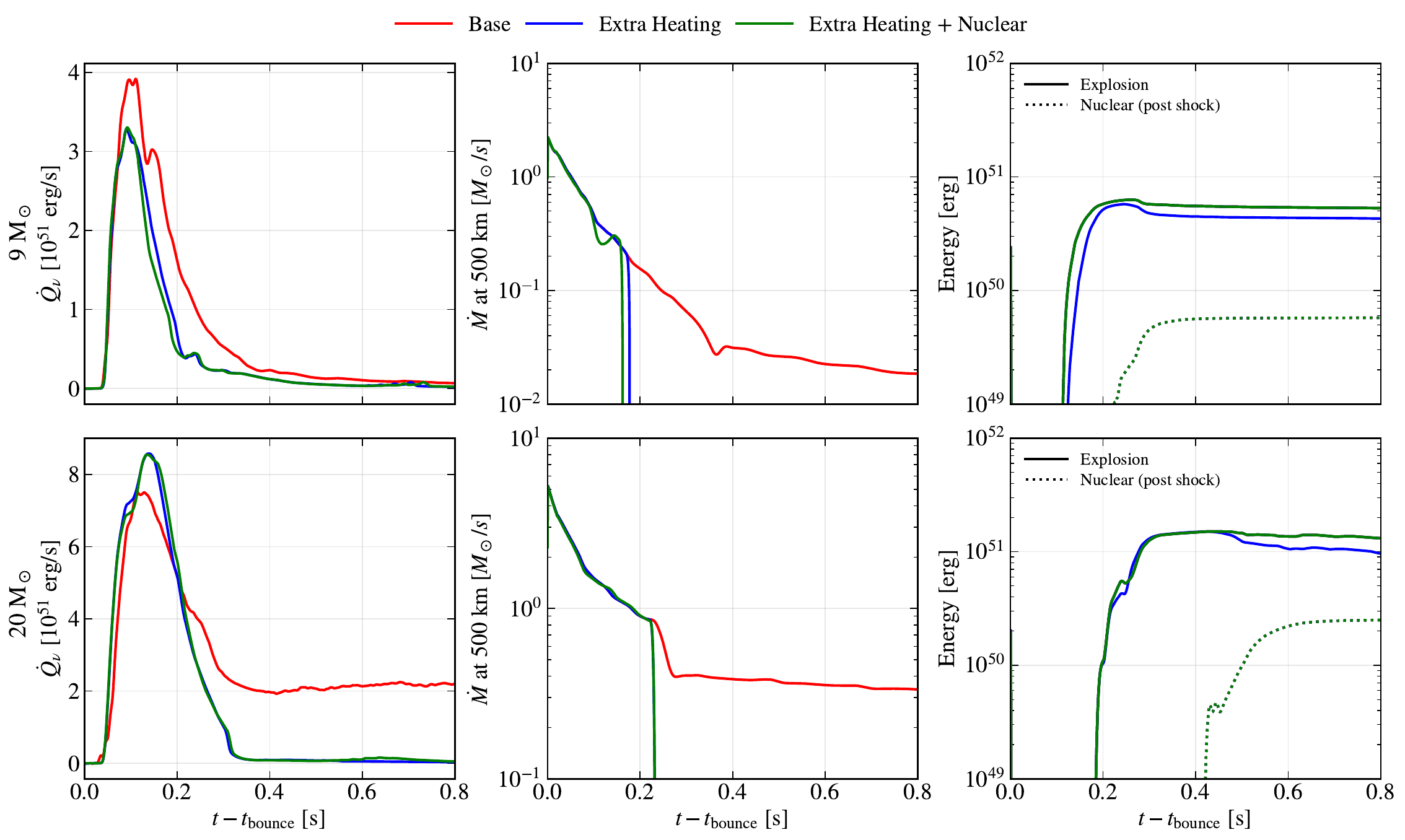}
	\caption{
		Time evolution of the net neutrino heating rate in the gain region (\emph{left}), mass accretion rate at 500~km (\emph{middle}), and diagnostic explosion energy (\emph{right}) in core-collapse supernova simulations of 9~M$_\odot$ (\emph{top}) and 20~M$_\odot$ (\emph{bottom}) progenitors. 
		In the exploding models, both the net neutrino heating rate and the mass accretion rate decrease significantly once shock revival is achieved. 
		These quantities exhibit similar behavior with and without nuclear burning, indicating that neutrino heating dominates the early explosion dynamics. 
		However, nuclear burning in the post-shock region contributes an additional $\mathcal{O}(10^{49{-}50})~\mathrm{erg}$ (dotted lines), corresponding to about $\mathcal{O}\left(10\right)~\%$ of the explosion energy. 
		This contribution enhances the final explosion energy and leads to faster shock expansion (not shown).
	}
	\label{fig:nuc_ccsn_accretion_erg_comparison}
\end{figure*}

To further quantify the role of nuclear burning in the explosion dynamics, Figure~\ref{fig:nuc_ccsn_accretion_erg_comparison} compares the time evolution of neutrino heating rate, mass accretion rate, and diagnostic explosion energy. 
In both cases, the onset of explosion is primarily controlled by neutrino heating, as reflected by the similar decline in heating rate and accretion rate once shock revival occurs, regardless of whether nuclear burning is included. 
Nevertheless, burning behind the shock contributes an additional $\sim$10\% to the explosion energy, amounting to $\mathcal{O}(10^{49{-}50})$~erg. 
Although subdominant compared to neutrino heating, this contribution is non-negligible: it strengthens the explosion energetics, accelerates shock expansion, and directly alters the composition of the ejecta. 

These 1D CCSN simulations demonstrate that \texttt{Gmunu} can robustly evolve the coupled system of hydrodynamics, gravity, neutrino transport, and nuclear burning under challenging conditions. 
While baseline models without enhanced neutrino heating fail to explode, consistent with previous 1D results, the inclusion of an artificial heating factor revives the shock. 
Incorporating nuclear burning further modifies the post-shock composition and dynamics, highlighting the role of explosive nucleosynthesis in shaping the ejecta. 
Together, these tests verify that the nuclear burning module integrates consistently with the microphysics and dynamical solvers, providing a foundation for future multidimensional, self-consistent CCSN studies with \texttt{Gmunu}. 
Looking ahead, these capabilities will enable detailed investigations of nucleosynthesis yields, neutrino signals, and gravitational-wave signatures from core-collapse supernovae in more realistic multidimensional settings.

\section{\label{sec:conclusions}Conclusions}
We have presented the implementation of nuclear reaction networks in the general-relativistic radiation hydrodynamics code \texttt{Gmunu}.
The new module has been verified through a suite of benchmark problems, including conserved-to-primitive tests with a stellar equation of state, one-zone silicon burning, and Newtonian shock-tube, acoustic pulse tests and detonation fronts of Type Ia supernovae.
These results confirm that the coupling between the nuclear network, the equation of state, and the hydrodynamics solver is robust, conserving both the electron fraction and the total mass fraction to machine precision.

We further demonstrated the application of this framework to one-dimensional core-collapse supernova simulations.
The results reproduce the expected non-exploding behavior of baseline 1D models and confirm that artificially enhanced neutrino heating can trigger shock revival.
When nuclear burning is included, the post-shock composition evolves dynamically, with silicon and oxygen layers progressively converted to iron-group nuclei as the shock expands and cools.
This highlights the role of explosive burning in shaping ejecta composition and demonstrates that reduced nuclear networks can be stably integrated in fully coupled, general-relativistic, radiation-hydrodynamic simulations.

Importantly, to the best of our knowledge, \texttt{Gmunu} is the first code that combines general relativity, M1 neutrino transport, magnetohydrodynamics, and in-situ nuclear burning within a single framework.
Although the current networks are limited in size, they provide a valuable bridge between simplified hydrodynamic models and large-scale post-processing calculations, offering more realistic composition and thermal profiles to seed follow-up nucleosynthesis studies.

Looking ahead, this work establishes a robust foundation for exploring a wide range of astrophysical explosion scenarios.
Future developments will focus on:
(i) multidimensional CCSN simulations with self-consistent neutrino heating to capture convection, turbulence, and asphericity;
(ii) adaptive nuclear networks that expand dynamically depending on local conditions, eventually incorporating up-to-date libraries such as REACLIB and {\tt pynucastro}~\citep{pynucastro, pynucastro2, pynucastro_development_team_2025_17455462};
(iii) systematic studies of nucleosynthesis yields across progenitor models; and
(iv) connections to multi-messenger observables, including neutrinos, gravitational waves, and electromagnetic transients.
Beyond core-collapse supernovae, the framework is also directly applicable to other explosive phenomena where nuclear burning plays a central role, such as collapsars, white dwarf–neutron star mergers, and white dwarf–white dwarf mergers.
Together, these advances will allow \texttt{Gmunu} to become a comprehensive platform for probing the interplay of gravity, neutrino transport, magnetism, and nuclear burning in some of the universe’s most energetic events.

\clearpage

\begin{acknowledgments}
We thank Albino Perego and David Radice for valuable discussions on bridging NSE and non-NSE equations of state. 
We are also grateful to Gerard Nav\'o and Martin Obergaulinger for their insights regarding the initial setup of the core-collapse supernova tests. 
We further thank Peter Siu Hei Cheung for helpful general discussions. 
P.C.-K.C. acknowledges support from the National Science Foundation (NSF) under Grant PHY-2020275, awarded to the Network for Neutrinos, Nuclear Astrophysics, and Symmetries (N3AS).

The simulations in this work have been performed on the third UNH supercomputer Marvin, also known as Plasma, which is supported by NSF/Major Research Instrumentatio (MRI) program under grant number AGS-1919310. 
The simulations have also been performed on the Expanse cluster at San Diego Supercomputer Centre through allocation PHY230104 and PHY230129 from the Advanced Cyberinfrastructure Coordination Ecosystem: Services \& Support (ACCESS) program~\citep{10.1145/3569951.3597559}, which is supported by National Science Foundation grants \#2138259, \#2138286, \#2138307, \#2137603, and \#2138296. 
\end{acknowledgments}


%\vspace{5mm}
%\facilities{HST(STIS), Swift(XRT and UVOT), AAVSO, CTIO:1.3m,CTIO:1.5m,CXO}


\software{
{
The results of this work were produced by utilising \texttt{Gmunu}~\citep{2020CQGra..37n5015C, 2021MNRAS.508.2279C} and the energy-dependent M1 neutrino transport module~\citep{2023ApJS..267...38C}.
The tabulated neutrino interaction were provided with \texttt{NuLib}~\citep{2015ApJS..219...24O}.
The EoS in the non-NSE region is provided with \texttt{helmeos}~\citep{2000ApJS..126..501T}. 
The nuclear networks used in this work (i.e. the \texttt{aprox13})~\citep{1999ApJS..124..241T, 2000ApJS..129..377T} are publicly available at \url{https://cococubed.com/}.
The data of the simulations were post-processed and visualised with 
\texttt{yt}~\citep{2011ApJS..192....9T},
\texttt{NumPy}~\citep{harris2020array}, 
\texttt{pandas}~\citep{reback2020pandas, mckinney-proc-scipy-2010},
\texttt{SciPy}~\citep{2020SciPy-NMeth} and
\texttt{Matplotlib}~\citep{2007CSE.....9...90H, thomas_a_caswell_2023_7697899}.
}
}



% \appendix


\bibliography{references}{}

@ARTICLE{2020CQGra..37n5015C,
       author = {{Cheong}, Patrick Chi-Kit and {Lin}, Lap-Ming and {Li}, Tjonnie Guang Feng},
        title = "{Gmunu: toward multigrid based Einstein field equations solver for general-relativistic hydrodynamics simulations}",
      journal = {Classical and Quantum Gravity},
         year = 2020,
        month = jul,
       volume = {37},
       number = {14},
          eid = {145015},
        pages = {145015},
          doi = {10.1088/1361-6382/ab8e9c},
archivePrefix = {arXiv},
       eprint = {2001.05723},
 primaryClass = {gr-qc},
       adsurl = {https://ui.adsabs.harvard.edu/abs/2020CQGra..37n5015C}
}

@ARTICLE{2021MNRAS.508.2279C,
       author = {{Cheong}, Patrick Chi-Kit and {Lam}, Alan Tsz-Lok and {Ng}, Harry Ho-Yin and {Li}, Tjonnie Guang Feng},
        title = "{Gmunu: paralleled, grid-adaptive, general-relativistic magnetohydrodynamics in curvilinear geometries in dynamical space-times}",
      journal = {\mnras},
         year = 2021,
        month = dec,
       volume = {508},
       number = {2},
        pages = {2279-2301},
          doi = {10.1093/mnras/stab2606},
archivePrefix = {arXiv},
       eprint = {2012.07322},
 primaryClass = {astro-ph.IM},
       adsurl = {https://ui.adsabs.harvard.edu/abs/2021MNRAS.508.2279C}
}

@ARTICLE{2022ApJS..261...22C,
       author = {{Cheong}, Patrick Chi-Kit and {Pong}, David Yat Tung and {Yip}, Anson Ka Long and {Li}, Tjonnie Guang Feng},
        title = "{An Extension of Gmunu: General-relativistic Resistive Magnetohydrodynamics Based on Staggered-meshed Constrained Transport with Elliptic Cleaning}",
      journal = {\apjs},
         year = 2022,
        month = aug,
       volume = {261},
       number = {2},
          eid = {22},
        pages = {22},
          doi = {10.3847/1538-4365/ac6cec},
archivePrefix = {arXiv},
       eprint = {2110.03732},
 primaryClass = {astro-ph.IM},
       adsurl = {https://ui.adsabs.harvard.edu/abs/2022ApJS..261...22C}
}

@ARTICLE{2023ApJS..267...38C,
       author = {{Cheong}, Patrick Chi-Kit and {Ng}, Harry Ho-Yin and {Lam}, Alan Tsz-Lok and {Li}, Tjonnie Guang Feng},
        title = "{General-relativistic Radiation Transport Scheme in Gmunu. I. Implementation of Two-moment-based Multifrequency Radiative Transfer and Code Tests}",
      journal = {\apjs},
         year = 2023,
        month = aug,
       volume = {267},
       number = {2},
          eid = {38},
        pages = {38},
          doi = {10.3847/1538-4365/acd931},
archivePrefix = {arXiv},
       eprint = {2303.03261},
 primaryClass = {astro-ph.IM},
       adsurl = {https://ui.adsabs.harvard.edu/abs/2023ApJS..267...38C}
}

@ARTICLE{2024ApJS..272....9N,
       author = {{Ng}, Harry Ho-Yin and {Cheong}, Patrick Chi-Kit and {Lam}, Alan Tsz-Lok and {Li}, Tjonnie Guang Feng},
        title = "{General-relativistic Radiation Transport Scheme in Gmunu. II. Implementation of Novel Microphysical Library for Neutrino Radiation{\textemdash}Weakhub}",
      journal = {\apjs},
         year = 2024,
        month = may,
       volume = {272},
       number = {1},
          eid = {9},
        pages = {9},
          doi = {10.3847/1538-4365/ad2fbd},
archivePrefix = {arXiv},
       eprint = {2309.03526},
 primaryClass = {astro-ph.HE},
       adsurl = {https://ui.adsabs.harvard.edu/abs/2024ApJS..272....9N}
}

@ARTICLE{2015ApJS..219...24O,
       author = {{O'Connor}, Evan},
        title = "{An Open-source Neutrino Radiation Hydrodynamics Code for Core-collapse Supernovae}",
      journal = {\apjs},
         year = 2015,
        month = aug,
       volume = {219},
       number = {2},
          eid = {24},
        pages = {24},
          doi = {10.1088/0067-0049/219/2/24},
archivePrefix = {arXiv},
       eprint = {1411.7058},
 primaryClass = {astro-ph.HE},
       adsurl = {https://ui.adsabs.harvard.edu/abs/2015ApJS..219...24O}
}

@ARTICLE{2000ApJS..126..501T,
       author = {{Timmes}, F.~X. and {Swesty}, F. Douglas},
        title = "{The Accuracy, Consistency, and Speed of an Electron-Positron Equation of State Based on Table Interpolation of the Helmholtz Free Energy}",
      journal = {\apjs},
         year = 2000,
        month = feb,
       volume = {126},
       number = {2},
        pages = {501-516},
          doi = {10.1086/313304},
       adsurl = {https://ui.adsabs.harvard.edu/abs/2000ApJS..126..501T}
}

@ARTICLE{1999ApJS..124..241T,
       author = {{Timmes}, F.~X.},
        title = "{Integration of Nuclear Reaction Networks for Stellar Hydrodynamics}",
      journal = {\apjs},
         year = 1999,
        month = sep,
       volume = {124},
       number = {1},
        pages = {241-263},
          doi = {10.1086/313257},
       adsurl = {https://ui.adsabs.harvard.edu/abs/1999ApJS..124..241T}
}

@ARTICLE{2000ApJS..129..377T,
       author = {{Timmes}, F.~X. and {Hoffman}, R.~D. and {Woosley}, S.~E.},
        title = "{An Inexpensive Nuclear Energy Generation Network for Stellar Hydrodynamics}",
      journal = {\apjs},
         year = 2000,
        month = jul,
       volume = {129},
       number = {1},
        pages = {377-398},
          doi = {10.1086/313407},
       adsurl = {https://ui.adsabs.harvard.edu/abs/2000ApJS..129..377T}
}

@ARTICLE{2007CSE.....9...90H,
       author = {{Hunter}, John D.},
        title = "{Matplotlib: A 2D Graphics Environment}",
      journal = {Computing in Science and Engineering},
         year = 2007,
        month = may,
       volume = {9},
       number = {3},
        pages = {90-95},
          doi = {10.1109/MCSE.2007.55},
       adsurl = {https://ui.adsabs.harvard.edu/abs/2007CSE.....9...90H}
}

@software{thomas_a_caswell_2023_7697899,
  author       = {Thomas A Caswell and
                  Antony Lee and
                  Elliott Sales de Andrade and
                  Michael Droettboom and
                  Tim Hoffmann and
                  Jody Klymak and
                  John Hunter and
                  Eric Firing and
                  David Stansby and
                  Nelle Varoquaux and
                  Jens Hedegaard Nielsen and
                  Benjamin Root and
                  Ryan May and
                  Oscar Gustafsson and
                  Phil Elson and
                  Jouni K. Seppänen and
                  Jae-Joon Lee and
                  Darren Dale and
                  hannah and
                  Damon McDougall and
                  Andrew Straw and
                  Paul Hobson and
                  Kyle Sunden and
                  Greg Lucas and
                  Christoph Gohlke and
                  Adrien F. Vincent and
                  Tony S Yu and
                  Eric Ma and
                  Steven Silvester and
                  Charlie Moad},
  title        = {matplotlib/matplotlib: REL: v3.7.1},
  month        = mar,
  year         = 2023,
  publisher    = {Zenodo},
  version      = {v3.7.1},
  doi          = {10.5281/zenodo.7697899},
  url          = {https://doi.org/10.5281/zenodo.7697899}
}

@Article{harris2020array,
 title         = {Array programming with {NumPy}},
 author        = {Charles R. Harris and K. Jarrod Millman and St{\'{e}}fan J.
                 van der Walt and Ralf Gommers and Pauli Virtanen and David
                 Cournapeau and Eric Wieser and Julian Taylor and Sebastian
                 Berg and Nathaniel J. Smith and Robert Kern and Matti Picus
                 and Stephan Hoyer and Marten H. van Kerkwijk and Matthew
                 Brett and Allan Haldane and Jaime Fern{\'{a}}ndez del
                 R{\'{i}}o and Mark Wiebe and Pearu Peterson and Pierre
                 G{\'{e}}rard-Marchant and Kevin Sheppard and Tyler Reddy and
                 Warren Weckesser and Hameer Abbasi and Christoph Gohlke and
                 Travis E. Oliphant},
 year          = {2020},
 month         = sep,
 journal       = {Nature},
 volume        = {585},
 number        = {7825},
 pages         = {357--362},
 doi           = {10.1038/s41586-020-2649-2},
 publisher     = {Springer Science and Business Media {LLC}},
 url           = {https://doi.org/10.1038/s41586-020-2649-2}
}

@software{reback2020pandas,
    author       = {The pandas development team},
    title        = {pandas-dev/pandas: Pandas},
    month        = feb,
    year         = 2020,
    publisher    = {Zenodo},
    version      = {latest},
    doi          = {10.5281/zenodo.3509134},
    url          = {https://doi.org/10.5281/zenodo.3509134}
}

@InProceedings{ mckinney-proc-scipy-2010,
  author    = { {W}es {M}c{K}inney },
  title     = { {D}ata {S}tructures for {S}tatistical {C}omputing in {P}ython },
  booktitle = { {P}roceedings of the 9th {P}ython in {S}cience {C}onference },
  pages     = { 56 - 61 },
  year      = { 2010 },
  editor    = { {S}t\'efan van der {W}alt and {J}arrod {M}illman },
  doi       = { 10.25080/Majora-92bf1922-00a }
}

@ARTICLE{2020SciPy-NMeth,
  author  = {Virtanen, Pauli and Gommers, Ralf and Oliphant, Travis E. and
            Haberland, Matt and Reddy, Tyler and Cournapeau, David and
            Burovski, Evgeni and Peterson, Pearu and Weckesser, Warren and
            Bright, Jonathan and {van der Walt}, St{\'e}fan J. and
            Brett, Matthew and Wilson, Joshua and Millman, K. Jarrod and
            Mayorov, Nikolay and Nelson, Andrew R. J. and Jones, Eric and
            Kern, Robert and Larson, Eric and Carey, C J and
            Polat, {\.I}lhan and Feng, Yu and Moore, Eric W. and
            {VanderPlas}, Jake and Laxalde, Denis and Perktold, Josef and
            Cimrman, Robert and Henriksen, Ian and Quintero, E. A. and
            Harris, Charles R. and Archibald, Anne M. and
            Ribeiro, Ant{\^o}nio H. and Pedregosa, Fabian and
            {van Mulbregt}, Paul and {SciPy 1.0 Contributors}},
  title   = {{{SciPy} 1.0: Fundamental Algorithms for Scientific
            Computing in Python}},
  journal = {Nature Methods},
  year    = {2020},
  volume  = {17},
  pages   = {261--272},
  adsurl  = {https://rdcu.be/b08Wh},
  doi     = {10.1038/s41592-019-0686-2},
}

@ARTICLE{2011ApJS..192....9T,
   author = {{Turk}, M.~J. and {Smith}, B.~D. and {Oishi}, J.~S. and {Skory}, S. and
     {Skillman}, S.~W. and {Abel}, T. and {Norman}, M.~L.},
    title = "{yt: A Multi-code Analysis Toolkit for Astrophysical Simulation Data}",
  journal = {The Astrophysical Journal Supplement Series},
archivePrefix = "arXiv",
   eprint = {1011.3514},
 primaryClass = "astro-ph.IM",
     year = 2011,
    month = jan,
   volume = 192,
      eid = {9},
    pages = {9},
      doi = {10.1088/0067-0049/192/1/9},
   adsurl = {https://ui.adsabs.harvard.edu/abs/2011ApJS..192....9T}
}

@inproceedings{10.1145/3569951.3597559,
author = {Boerner, Timothy J. and Deems, Stephen and Furlani, Thomas R. and Knuth, Shelley L. and Towns, John},
title = {ACCESS: Advancing Innovation: NSF’s Advanced Cyberinfrastructure Coordination Ecosystem: Services \& Support},
year = {2023},
isbn = {9781450399852},
publisher = {Association for Computing Machinery},
address = {New York, NY, USA},
url = {https://doi.org/10.1145/3569951.3597559},
doi = {10.1145/3569951.3597559},
booktitle = {Practice and Experience in Advanced Research Computing},
pages = {173–176},
numpages = {4},
location = {Portland, OR, USA},
series = {PEARC '23}
}

@ARTICLE{2015ApJS..216...31Z,
       author = {{Zingale}, Michael and {Katz}, Max P.},
        title = "{On the Piecewise Parabolic Method for Compressible Flow With Stellar Equations of State}",
      journal = {\apjs},
         year = 2015,
        month = feb,
       volume = {216},
       number = {2},
          eid = {31},
        pages = {31},
          doi = {10.1088/0067-0049/216/2/31},
archivePrefix = {arXiv},
       eprint = {1501.01923},
 primaryClass = {astro-ph.IM},
       adsurl = {https://ui.adsabs.harvard.edu/abs/2015ApJS..216...31Z}
}

@ARTICLE{2019ApJ...886..105Z,
       author = {{Zingale}, M. and {Katz}, M.~P. and {Bell}, J.~B. and {Minion}, M.~L. and {Nonaka}, A.~J. and {Zhang}, W.},
        title = "{Improved Coupling of Hydrodynamics and Nuclear Reactions via Spectral Deferred Corrections}",
      journal = {\apj},
         year = 2019,
        month = dec,
       volume = {886},
       number = {2},
          eid = {105},
        pages = {105},
          doi = {10.3847/1538-4357/ab4e1d},
archivePrefix = {arXiv},
       eprint = {1908.03661},
 primaryClass = {physics.comp-ph},
       adsurl = {https://ui.adsabs.harvard.edu/abs/2019ApJ...886..105Z}
}

@ARTICLE{2025ApJ...981...63H,
       author = {{Hasenour}, Dillon L. and {Duffell}, Paul C.},
        title = "{Quantifying Advantages of a Moving Mesh in Nuclear Hydrodynamics}",
      journal = {\apj},
         year = 2025,
        month = mar,
       volume = {981},
       number = {1},
          eid = {63},
        pages = {63},
          doi = {10.3847/1538-4357/adaeb0},
archivePrefix = {arXiv},
       eprint = {2502.02693},
 primaryClass = {astro-ph.HE},
       adsurl = {https://ui.adsabs.harvard.edu/abs/2025ApJ...981...63H}
}

@article{pareschi2005implicit,
  title={Implicit--explicit Runge--Kutta schemes and applications to hyperbolic systems with relaxation},
  author={Pareschi, Lorenzo and Russo, Giovanni},
  journal={Journal of Scientific computing},
  volume={25},
  number={1},
  pages={129--155},
  year={2005},
  publisher={Springer}
}

@ARTICLE{1974JCoPh..14..361V,
       author = {{van Leer}, Bram},
        title = "{Towards the Ultimate Conservation Difference Scheme. II. Monotonicity and Conservation Combined in a Second-Order Scheme}",
      journal = {Journal of Computational Physics},
         year = 1974,
        month = mar,
       volume = {14},
       number = {4},
        pages = {361-370},
          doi = {10.1016/0021-9991(74)90019-9},
       adsurl = {https://ui.adsabs.harvard.edu/abs/1974JCoPh..14..361V}
}

@ARTICLE{2002A&A...396..361R,
       author = {{Rampp}, M. and {Janka}, H. -T.},
        title = "{Radiation hydrodynamics with neutrinos. Variable Eddington factor method for core-collapse supernova simulations}",
      journal = {\aap},
         year = 2002,
        month = dec,
       volume = {396},
        pages = {361-392},
          doi = {10.1051/0004-6361:20021398},
archivePrefix = {arXiv},
       eprint = {astro-ph/0203101},
 primaryClass = {astro-ph},
       adsurl = {https://ui.adsabs.harvard.edu/abs/2002A&A...396..361R}
}

@ARTICLE{2006A&A...447.1049B,
       author = {{Buras}, R. and {Rampp}, M. and {Janka}, H. -Th. and {Kifonidis}, K.},
        title = "{Two-dimensional hydrodynamic core-collapse supernova simulations with spectral neutrino transport. I. Numerical method and results for a 15 M{\ensuremath{\odot}} star}",
      journal = {\aap},
         year = 2006,
        month = mar,
       volume = {447},
       number = {3},
        pages = {1049-1092},
          doi = {10.1051/0004-6361:20053783},
archivePrefix = {arXiv},
       eprint = {astro-ph/0507135},
 primaryClass = {astro-ph},
       adsurl = {https://ui.adsabs.harvard.edu/abs/2006A&A...447.1049B}
}

@ARTICLE{2023ApJ...951..112N,
       author = {{Nav{\'o}}, Gerard and {Reichert}, Moritz and {Obergaulinger}, Martin and {Arcones}, Almudena},
        title = "{Core-collapse Supernova Simulations with Reduced Nucleosynthesis Networks}",
      journal = {\apj},
         year = 2023,
        month = jul,
       volume = {951},
       number = {2},
          eid = {112},
        pages = {112},
          doi = {10.3847/1538-4357/acd640},
archivePrefix = {arXiv},
       eprint = {2210.11848},
 primaryClass = {astro-ph.HE},
       adsurl = {https://ui.adsabs.harvard.edu/abs/2023ApJ...951..112N}
}

@ARTICLE{2016ApJ...818..123B,
       author = {{Bruenn}, Stephen W. and {Lentz}, Eric J. and {Hix}, W. Raphael and {Mezzacappa}, Anthony and {Harris}, J. Austin and {Messer}, O.~E. Bronson and {Endeve}, Eirik and {Blondin}, John M. and {Chertkow}, Merek Austin and {Lingerfelt}, Eric J. and {Marronetti}, Pedro and {Yakunin}, Konstantin N.},
        title = "{The Development of Explosions in Axisymmetric Ab Initio Core-collapse Supernova Simulations of 12-25 M Stars}",
      journal = {\apj},
         year = 2016,
        month = feb,
       volume = {818},
       number = {2},
          eid = {123},
        pages = {123},
          doi = {10.3847/0004-637X/818/2/123},
archivePrefix = {arXiv},
       eprint = {1409.5779},
 primaryClass = {astro-ph.SR},
       adsurl = {https://ui.adsabs.harvard.edu/abs/2016ApJ...818..123B}
}

@ARTICLE{2020ApJS..248...11B,
       author = {{Bruenn}, Stephen W. and {Blondin}, John M. and {Hix}, W. Raphael and {Lentz}, Eric J. and {Messer}, O.~E. Bronson and {Mezzacappa}, Anthony and {Endeve}, Eirik and {Harris}, J. Austin and {Marronetti}, Pedro and {Budiardja}, Reuben D. and {Chertkow}, Merek A. and {Lee}, Ching-Tsai},
        title = "{CHIMERA: A Massively Parallel Code for Core-collapse Supernova Simulations}",
      journal = {\apjs},
         year = 2020,
        month = may,
       volume = {248},
       number = {1},
          eid = {11},
        pages = {11},
          doi = {10.3847/1538-4365/ab7aff},
archivePrefix = {arXiv},
       eprint = {1809.05608},
 primaryClass = {astro-ph.IM},
       adsurl = {https://ui.adsabs.harvard.edu/abs/2020ApJS..248...11B}
}

@ARTICLE{2015ApJ...806..275P,
       author = {{Perego}, A. and {Hempel}, M. and {Fr{\"o}hlich}, C. and {Ebinger}, K. and {Eichler}, M. and {Casanova}, J. and {Liebend{\"o}rfer}, M. and {Thielemann}, F. -K.},
        title = "{PUSHing Core-collapse Supernovae to Explosions in Spherical Symmetry I: the Model and the Case of SN 1987A}",
      journal = {\apj},
         year = 2015,
        month = jun,
       volume = {806},
       number = {2},
          eid = {275},
        pages = {275},
          doi = {10.1088/0004-637X/806/2/275},
archivePrefix = {arXiv},
       eprint = {1501.02845},
 primaryClass = {astro-ph.SR},
       adsurl = {https://ui.adsabs.harvard.edu/abs/2015ApJ...806..275P}
}

@ARTICLE{2010CQGra..27k4103O,
       author = {{O'Connor}, Evan and {Ott}, Christian D.},
        title = "{A new open-source code for spherically symmetric stellar collapse to neutron stars and black holes}",
      journal = {Classical and Quantum Gravity},
         year = 2010,
        month = jun,
       volume = {27},
       number = {11},
          eid = {114103},
        pages = {114103},
          doi = {10.1088/0264-9381/27/11/114103},
archivePrefix = {arXiv},
       eprint = {0912.2393},
 primaryClass = {astro-ph.HE},
       adsurl = {https://ui.adsabs.harvard.edu/abs/2010CQGra..27k4103O}
}

@ARTICLE{2025ApJ...981..119F,
       author = {{Fujibayashi}, Sho and {Jockel}, C{\'e}dric and {Kawaguchi}, Kyohei and {Sekiguchi}, Yuichiro and {Shibata}, Masaru},
        title = "{Powerful Explosions from the Collapse of Rotating Supermassive Stars}",
      journal = {\apj},
         year = 2025,
        month = mar,
       volume = {981},
       number = {2},
          eid = {119},
        pages = {119},
          doi = {10.3847/1538-4357/adb0b8},
archivePrefix = {arXiv},
       eprint = {2408.11572},
 primaryClass = {astro-ph.HE},
       adsurl = {https://ui.adsabs.harvard.edu/abs/2025ApJ...981..119F}
}

@ARTICLE{2012ApJ...749...37M,
       author = {{Montero}, Pedro J. and {Janka}, Hans-Thomas and {M{\"u}ller}, Ewald},
        title = "{Relativistic Collapse and Explosion of Rotating Supermassive Stars with Thermonuclear Effects}",
      journal = {\apj},
         year = 2012,
        month = apr,
       volume = {749},
       number = {1},
          eid = {37},
        pages = {37},
          doi = {10.1088/0004-637X/749/1/37},
archivePrefix = {arXiv},
       eprint = {1108.3090},
 primaryClass = {astro-ph.CO},
       adsurl = {https://ui.adsabs.harvard.edu/abs/2012ApJ...749...37M}
}

@ARTICLE{2017PhRvD..96h3016U,
       author = {{Uchida}, Haruki and {Shibata}, Masaru and {Yoshida}, Takashi and {Sekiguchi}, Yuichiro and {Umeda}, Hideyuki},
        title = "{Gravitational collapse of rotating supermassive stars including nuclear burning effects}",
      journal = {\prd},
         year = 2017,
        month = oct,
       volume = {96},
       number = {8},
          eid = {083016},
        pages = {083016},
          doi = {10.1103/PhysRevD.96.083016},
archivePrefix = {arXiv},
       eprint = {1704.00433},
 primaryClass = {astro-ph.HE},
       adsurl = {https://ui.adsabs.harvard.edu/abs/2017PhRvD..96h3016U}
}

@article{ASCHER1997151,
title = {Implicit-explicit Runge-Kutta methods for time-dependent partial differential equations},
journal = {Applied Numerical Mathematics},
volume = {25},
number = {2},
pages = {151-167},
year = {1997},
note = {Special Issue on Time Integration},
issn = {0168-9274},
doi = {https://doi.org/10.1016/S0168-9274(97)00056-1},
url = {https://www.sciencedirect.com/science/article/pii/S0168927497000561},
author = {Uri M. Ascher and Steven J. Ruuth and Raymond J. Spiteri}
}

@book{press1996numerical,
  title={Numerical recipes in Fortran 90},
  author={Press, William H and Teukolsky, Saul A and Vetterling, William T and Flannery, Brian P},
  volume={2},
  year={1996},
  publisher={Cambridge university press Cambridge}
}

@ARTICLE{2014ApJ...782...91N,
       author = {{Nakamura}, Ko and {Takiwaki}, Tomoya and {Kotake}, Kei and {Nishimura}, Nobuya},
        title = "{Revisiting Impacts of Nuclear Burning for Reviving Weak Shocks in Neutrino-driven Supernovae}",
      journal = {\apj},
         year = 2014,
        month = feb,
       volume = {782},
       number = {2},
          eid = {91},
        pages = {91},
          doi = {10.1088/0004-637X/782/2/91},
archivePrefix = {arXiv},
       eprint = {1207.5955},
 primaryClass = {astro-ph.HE},
       adsurl = {https://ui.adsabs.harvard.edu/abs/2014ApJ...782...91N}
}

@ARTICLE{2014PhRvD..89h4043M,
       author = {{Montero}, Pedro J. and {Baumgarte}, Thomas W. and {M{\"u}ller}, Ewald},
        title = "{General relativistic hydrodynamics in curvilinear coordinates}",
      journal = {\prd},
         year = 2014,
        month = apr,
       volume = {89},
       number = {8},
          eid = {084043},
        pages = {084043},
          doi = {10.1103/PhysRevD.89.084043},
archivePrefix = {arXiv},
       eprint = {1309.7808},
 primaryClass = {gr-qc},
       adsurl = {https://ui.adsabs.harvard.edu/abs/2014PhRvD..89h4043M}
}

@ARTICLE{2020PhRvD.101j4007M,
       author = {{Mewes}, Vassilios and {Zlochower}, Yosef and {Campanelli}, Manuela and
         {Baumgarte}, Thomas W. and {Etienne}, Zachariah B. and
         {Armengol}, Federico G. Lopez and {Cipolletta}, Federico},
        title = "{Numerical relativity in spherical coordinates: A new dynamical spacetime and general relativistic MHD evolution framework for the Einstein Toolkit}",
      journal = {\prd},
         year = 2020,
        month = may,
       volume = {101},
       number = {10},
          eid = {104007},
        pages = {104007},
          doi = {10.1103/PhysRevD.101.104007},
archivePrefix = {arXiv},
       eprint = {2002.06225},
 primaryClass = {gr-qc},
       adsurl = {https://ui.adsabs.harvard.edu/abs/2020PhRvD.101j4007M}
}

@ARTICLE{2020PhRvD.102j4001B,
       author = {{Baumgarte}, Thomas W. and {Shapiro}, Stuart L.},
        title = "{Relativistic radiation hydrodynamics in a reference-metric formulation}",
      journal = {\prd},
         year = 2020,
        month = nov,
       volume = {102},
       number = {10},
          eid = {104001},
        pages = {104001},
          doi = {10.1103/PhysRevD.102.104001},
archivePrefix = {arXiv},
       eprint = {2009.08990},
 primaryClass = {gr-qc},
       adsurl = {https://ui.adsabs.harvard.edu/abs/2020PhRvD.102j4001B}
}

@ARTICLE{2024ApJ...975..116C,
       author = {{Cheong}, Patrick Chi-Kit and {Foucart}, Francois and {Duez}, Matthew D. and {Offermans}, Arthur and {Muhammed}, Nishad and {Chawhan}, Pavan},
        title = "{Energy-dependent and Energy-integrated Two-moment General-relativistic Neutrino Transport Simulations of a Hypermassive Neutron Star}",
      journal = {\apj},
         year = 2024,
        month = nov,
       volume = {975},
       number = {1},
          eid = {116},
        pages = {116},
          doi = {10.3847/1538-4357/ad7825},
archivePrefix = {arXiv},
       eprint = {2407.16017},
 primaryClass = {astro-ph.HE},
       adsurl = {https://ui.adsabs.harvard.edu/abs/2024ApJ...975..116C}
}

@ARTICLE{2015ApJ...808L..21C,
       author = {{Couch}, Sean M. and {Chatzopoulos}, Emmanouil and {Arnett}, W. David and {Timmes}, F.~X.},
        title = "{The Three-dimensional Evolution to Core Collapse of a Massive Star}",
      journal = {\apjl},
         year = 2015,
        month = jul,
       volume = {808},
       number = {1},
          eid = {L21},
        pages = {L21},
          doi = {10.1088/2041-8205/808/1/L21},
archivePrefix = {arXiv},
       eprint = {1503.02199},
 primaryClass = {astro-ph.HE},
       adsurl = {https://ui.adsabs.harvard.edu/abs/2015ApJ...808L..21C}
}

@article{doi:10.1137/0705041,
author = {Strang, Gilbert},
title = {On the Construction and Comparison of Difference Schemes},
journal = {SIAM Journal on Numerical Analysis},
volume = {5},
number = {3},
pages = {506-517},
year = {1968},
doi = {10.1137/0705041},
URL = {https://doi.org/10.1137/0705041},
eprint = {https://doi.org/10.1137/0705041}
}

@ARTICLE{2017ApJS..233...18L,
       author = {{Lippuner}, Jonas and {Roberts}, Luke F.},
        title = "{SkyNet: A Modular Nuclear Reaction Network Library}",
      journal = {\apjs},
         year = 2017,
        month = dec,
       volume = {233},
       number = {2},
          eid = {18},
        pages = {18},
          doi = {10.3847/1538-4365/aa94cb},
archivePrefix = {arXiv},
       eprint = {1706.06198},
 primaryClass = {astro-ph.HE},
       adsurl = {https://ui.adsabs.harvard.edu/abs/2017ApJS..233...18L}
}

@ARTICLE{1999JCoAM.109..321H, 
       author = {{Hix}, W.~R. and {Thielemann}, F. -K.},
        title = "{Computational methods for nucleosynthesis and nuclear energy generation.}",
      journal = {Journal of Computational and Applied Mathematics},
         year = 1999,
        month = sep,
       volume = {109},
       number = {1},
        pages = {321-351},
          doi = {10.48550/arXiv.astro-ph/9906478},
archivePrefix = {arXiv},
       eprint = {astro-ph/9906478},
 primaryClass = {astro-ph},
       adsurl = {https://ui.adsabs.harvard.edu/abs/1999JCoAM.109..321H}
}

@ARTICLE{1986A&A...162..103M,
       author = {{Mueller}, E.},
        title = "{Nuclear-reaction networks and stellar evolution codes - The coupling of composition changes and energy release in explosive nuclear burning}",
      journal = {\aap},
         year = 1986,
        month = jul,
       volume = {162},
       number = {1-2},
        pages = {103-108},
       adsurl = {https://ui.adsabs.harvard.edu/abs/1986A&A...162..103M}
}

@ARTICLE{2007ApJ...656..313C,
       author = {{Calder}, A.~C. and {Townsley}, D.~M. and {Seitenzahl}, I.~R. and {Peng}, F. and {Messer}, O.~E.~B. and {Vladimirova}, N. and {Brown}, E.~F. and {Truran}, J.~W. and {Lamb}, D.~Q.},
        title = "{Capturing the Fire: Flame Energetics and Neutronization for Type Ia Supernova Simulations}",
      journal = {\apj},
         year = 2007,
        month = feb,
       volume = {656},
       number = {1},
        pages = {313-332},
          doi = {10.1086/510709},
archivePrefix = {arXiv},
       eprint = {astro-ph/0611009},
 primaryClass = {astro-ph},
       adsurl = {https://ui.adsabs.harvard.edu/abs/2007ApJ...656..313C}
}

@ARTICLE{2009ADNDT..95...96S,
       author = {{Seitenzahl}, Ivo R. and {Townsley}, Dean M. and {Peng}, Fang and {Truran}, James W.},
        title = "{Nuclear statistical equilibrium for Type Ia supernova simulations}",
      journal = {Atomic Data and Nuclear Data Tables},
         year = 2009,
        month = jan,
       volume = {95},
       number = {1},
        pages = {96-114},
          doi = {10.1016/j.adt.2008.08.001},
       adsurl = {https://ui.adsabs.harvard.edu/abs/2009ADNDT..95...96S}
}

@ARTICLE{2008ApJ...685L.129S,
       author = {{Seitenzahl}, I.~R. and {Timmes}, F.~X. and {Marin-Lafl{\`e}che}, A. and {Brown}, E. and {Magkotsios}, G. and {Truran}, J.},
        title = "{Proton-rich Nuclear Statistical Equilibrium}",
      journal = {\apjl},
         year = 2008,
        month = oct,
       volume = {685},
       number = {2},
        pages = {L129},
          doi = {10.1086/592501},
archivePrefix = {arXiv},
       eprint = {0808.2033},
 primaryClass = {astro-ph},
       adsurl = {https://ui.adsabs.harvard.edu/abs/2008ApJ...685L.129S}
}

@ARTICLE{2000ADNDT..75....1R,
       author = {{Rauscher}, Thomas and {Thielemann}, Friedrich-Karl},
        title = "{Astrophysical Reaction Rates From Statistical Model Calculations}",
      journal = {Atomic Data and Nuclear Data Tables},
         year = 2000,
        month = may,
       volume = {75},
       number = {1-2},
        pages = {1-351},
          doi = {10.1006/adnd.2000.0834},
archivePrefix = {arXiv},
       eprint = {astro-ph/0004059},
 primaryClass = {astro-ph},
       adsurl = {https://ui.adsabs.harvard.edu/abs/2000ADNDT..75....1R}
}

@ARTICLE{1998PhRvE..58.4941C,
       author = {{Chabrier}, Gilles and {Potekhin}, Alexander Y.},
        title = "{Equation of state of fully ionized electron-ion plasmas}",
      journal = {\pre},
         year = 1998,
        month = oct,
       volume = {58},
       number = {4},
        pages = {4941-4949},
          doi = {10.1103/PhysRevE.58.4941},
archivePrefix = {arXiv},
       eprint = {physics/9807042},
 primaryClass = {physics.plasm-ph},
       adsurl = {https://ui.adsabs.harvard.edu/abs/1998PhRvE..58.4941C}
}

@ARTICLE{2016ApJ...818..124E,
       author = {{Ertl}, T. and {Janka}, H. -Th. and {Woosley}, S.~E. and {Sukhbold}, T. and {Ugliano}, M.},
        title = "{A Two-parameter Criterion for Classifying the Explodability of Massive Stars by the Neutrino-driven Mechanism}",
      journal = {\apj},
         year = 2016,
        month = feb,
       volume = {818},
       number = {2},
          eid = {124},
        pages = {124},
          doi = {10.3847/0004-637X/818/2/124},
archivePrefix = {arXiv},
       eprint = {1503.07522},
 primaryClass = {astro-ph.SR},
       adsurl = {https://ui.adsabs.harvard.edu/abs/2016ApJ...818..124E}
}

@ARTICLE{2021ApJ...912...29B,
       author = {{Boccioli}, Luca and {Mathews}, Grant J. and {O'Connor}, Evan P.},
        title = "{General Relativistic Neutrino-driven Turbulence in One-dimensional Core-collapse Supernovae}",
      journal = {\apj},
         year = 2021,
        month = may,
       volume = {912},
       number = {1},
          eid = {29},
        pages = {29},
          doi = {10.3847/1538-4357/abe767},
archivePrefix = {arXiv},
       eprint = {2102.06767},
 primaryClass = {astro-ph.HE},
       adsurl = {https://ui.adsabs.harvard.edu/abs/2021ApJ...912...29B}
}

@ARTICLE{1993A&A...273..106B,
       author = {{Blinnikov}, S.~I. and {Bartunov}, O.~S.},
        title = "{Non-equilibrium radiative transfer in supernova theory : models of linear type II supernovae.}",
      journal = {\aap},
         year = 1993,
        month = jun,
       volume = {273},
        pages = {106-122},
       adsurl = {https://ui.adsabs.harvard.edu/abs/1993A&A...273..106B}
}

@ARTICLE{1995ApJS..101..181W,
       author = {{Woosley}, S.~E. and {Weaver}, Thomas A.},
        title = "{The Evolution and Explosion of Massive Stars. II. Explosive Hydrodynamics and Nucleosynthesis}",
      journal = {\apjs},
         year = 1995,
        month = nov,
       volume = {101},
        pages = {181},
          doi = {10.1086/192237},
       adsurl = {https://ui.adsabs.harvard.edu/abs/1995ApJS..101..181W}
}

@ARTICLE{2012ApJ...757...69U,
       author = {{Ugliano}, Marcella and {Janka}, Hans-Thomas and {Marek}, Andreas and {Arcones}, Almudena},
        title = "{Progenitor-explosion Connection and Remnant Birth Masses for Neutrino-driven Supernovae of Iron-core Progenitors}",
      journal = {\apj},
         year = 2012,
        month = sep,
       volume = {757},
       number = {1},
          eid = {69},
        pages = {69},
          doi = {10.1088/0004-637X/757/1/69},
archivePrefix = {arXiv},
       eprint = {1205.3657},
 primaryClass = {astro-ph.SR},
       adsurl = {https://ui.adsabs.harvard.edu/abs/2012ApJ...757...69U}
}

@ARTICLE{2020ApJ...890..127C,
       author = {{Couch}, Sean M. and {Warren}, MacKenzie L. and {O'Connor}, Evan P.},
        title = "{Simulating Turbulence-aided Neutrino-driven Core-collapse Supernova Explosions in One Dimension}",
      journal = {\apj},
         year = 2020,
        month = feb,
       volume = {890},
       number = {2},
          eid = {127},
        pages = {127},
          doi = {10.3847/1538-4357/ab609e},
archivePrefix = {arXiv},
       eprint = {1902.01340},
 primaryClass = {astro-ph.HE},
       adsurl = {https://ui.adsabs.harvard.edu/abs/2020ApJ...890..127C}
}

@ARTICLE{2007PhR...442..269W,
       author = {{Woosley}, S.~E. and {Heger}, A.},
        title = "{Nucleosynthesis and remnants in massive stars of solar metallicity}",
      journal = {\physrep},
         year = 2007,
        month = apr,
       volume = {442},
       number = {1-6},
        pages = {269-283},
          doi = {10.1016/j.physrep.2007.02.009},
archivePrefix = {arXiv},
       eprint = {astro-ph/0702176},
 primaryClass = {astro-ph},
       adsurl = {https://ui.adsabs.harvard.edu/abs/2007PhR...442..269W}
}

@ARTICLE{2013ApJ...774...17S,
       author = {{Steiner}, A.~W. and {Hempel}, M. and {Fischer}, T.},
        title = "{Core-collapse Supernova Equations of State Based on Neutron Star Observations}",
      journal = {\apj},
         year = 2013,
        month = sep,
       volume = {774},
       number = {1},
          eid = {17},
        pages = {17},
          doi = {10.1088/0004-637X/774/1/17},
archivePrefix = {arXiv},
       eprint = {1207.2184},
 primaryClass = {astro-ph.SR},
       adsurl = {https://ui.adsabs.harvard.edu/abs/2013ApJ...774...17S}
}

@ARTICLE{2018JPhG...45j4001O,
       author = {{O'Connor}, Evan and {Bollig}, Robert and {Burrows}, Adam and {Couch}, Sean and {Fischer}, Tobias and {Janka}, Hans-Thomas and {Kotake}, Kei and {Lentz}, Eric J. and {Liebend{\"o}rfer}, Matthias and {Messer}, O.~E. Bronson and {Mezzacappa}, Anthony and {Takiwaki}, Tomoya and {Vartanyan}, David},
        title = "{Global comparison of core-collapse supernova simulations in spherical symmetry}",
      journal = {Journal of Physics G Nuclear Physics},
         year = 2018,
        month = oct,
       volume = {45},
       number = {10},
        pages = {104001},
          doi = {10.1088/1361-6471/aadeae},
archivePrefix = {arXiv},
       eprint = {1806.04175},
 primaryClass = {astro-ph.HE},
       adsurl = {https://ui.adsabs.harvard.edu/abs/2018JPhG...45j4001O}
}

@ARTICLE{2023MNRAS.524.4109G,
       author = {{Gogilashvili}, Mariam and {Murphy}, Jeremiah W. and {O'Connor}, Evan P.},
        title = "{The force explosion condition is consistent with spherically symmetric CCSN explosions}",
      journal = {\mnras},
         year = 2023,
        month = sep,
       volume = {524},
       number = {3},
        pages = {4109-4115},
          doi = {10.1093/mnras/stad2155},
archivePrefix = {arXiv},
       eprint = {2302.04890},
 primaryClass = {astro-ph.HE},
       adsurl = {https://ui.adsabs.harvard.edu/abs/2023MNRAS.524.4109G}
}

@ARTICLE{2017ApJ...843....2H,
       author = {{Harris}, J. Austin and {Hix}, W. Raphael and {Chertkow}, Merek A. and {Lee}, C.~T. and {Lentz}, Eric J. and {Messer}, O.~E. Bronson},
        title = "{Implications for Post-processing Nucleosynthesis of Core-collapse Supernova Models with Lagrangian Particles}",
      journal = {\apj},
         year = 2017,
        month = jul,
       volume = {843},
       number = {1},
          eid = {2},
        pages = {2},
          doi = {10.3847/1538-4357/aa76de},
archivePrefix = {arXiv},
       eprint = {1701.08876},
 primaryClass = {astro-ph.SR},
       adsurl = {https://ui.adsabs.harvard.edu/abs/2017ApJ...843....2H}
}

@ARTICLE{2021PhRvD.103b3018K,
       author = {{Kastaun}, Wolfgang and {Kalinani}, Jay Vijay and {Ciolfi}, Riccardo},
        title = "{Robust recovery of primitive variables in relativistic ideal magnetohydrodynamics}",
      journal = {\prd},
         year = 2021,
        month = jan,
       volume = {103},
       number = {2},
          eid = {023018},
        pages = {023018},
          doi = {10.1103/PhysRevD.103.023018},
archivePrefix = {arXiv},
       eprint = {2005.01821},
 primaryClass = {gr-qc},
       adsurl = {https://ui.adsabs.harvard.edu/abs/2021PhRvD.103b3018K}
}

@ARTICLE{2018ApJ...859...71S,
       author = {{Siegel}, Daniel M. and {M{\"o}sta}, Philipp and {Desai}, Dhruv and {Wu}, Samantha},
        title = "{Recovery Schemes for Primitive Variables in General-relativistic Magnetohydrodynamics}",
      journal = {\apj},
         year = 2018,
        month = may,
       volume = {859},
       number = {1},
          eid = {71},
        pages = {71},
          doi = {10.3847/1538-4357/aabcc5},
archivePrefix = {arXiv},
       eprint = {1712.07538},
 primaryClass = {astro-ph.HE},
       adsurl = {https://ui.adsabs.harvard.edu/abs/2018ApJ...859...71S}
}

@ARTICLE{1973ApJS...26..231W,
       author = {{Woosley}, S.~E. and {Arnett}, W. David and {Clayton}, Donald D.},
        title = "{The Explosive Burning of Oxygen and Silicon}",
      journal = {\apjs},
         year = 1973,
        month = nov,
       volume = {26},
        pages = {231},
          doi = {10.1086/190282},
       adsurl = {https://ui.adsabs.harvard.edu/abs/1973ApJS...26..231W}
}

@ARTICLE{1996ApJ...460..869H,
       author = {{Hix}, W. Raphael and {Thielemann}, Friedrich-Karl},
        title = "{Silicon Burning. I. Neutronization and the Physics of Quasi-Equilibrium}",
      journal = {\apj},
         year = 1996,
        month = apr,
       volume = {460},
        pages = {869},
          doi = {10.1086/177016},
archivePrefix = {arXiv},
       eprint = {astro-ph/9511088},
 primaryClass = {astro-ph},
       adsurl = {https://ui.adsabs.harvard.edu/abs/1996ApJ...460..869H}
}

@ARTICLE{2016ApJ...821...38S,
       author = {{Sukhbold}, Tuguldur and {Ertl}, T. and {Woosley}, S.~E. and {Brown}, Justin M. and {Janka}, H. -T.},
        title = "{Core-collapse Supernovae from 9 to 120 Solar Masses Based on Neutrino-powered Explosions}",
      journal = {\apj},
         year = 2016,
        month = apr,
       volume = {821},
       number = {1},
          eid = {38},
        pages = {38},
          doi = {10.3847/0004-637X/821/1/38},
archivePrefix = {arXiv},
       eprint = {1510.04643},
 primaryClass = {astro-ph.HE},
       adsurl = {https://ui.adsabs.harvard.edu/abs/2016ApJ...821...38S}
}

@ARTICLE{2017ApJ...850...43R,
       author = {{Radice}, David and {Burrows}, Adam and {Vartanyan}, David and {Skinner}, M. Aaron and {Dolence}, Joshua C.},
        title = "{Electron-capture and Low-mass Iron-core-collapse Supernovae: New Neutrino-radiation-hydrodynamics Simulations}",
      journal = {\apj},
         year = 2017,
        month = nov,
       volume = {850},
       number = {1},
          eid = {43},
        pages = {43},
          doi = {10.3847/1538-4357/aa92c5},
archivePrefix = {arXiv},
       eprint = {1702.03927},
 primaryClass = {astro-ph.HE},
       adsurl = {https://ui.adsabs.harvard.edu/abs/2017ApJ...850...43R}
}

@ARTICLE{1991NuPhA.535..331L,
       author = {{Lattimer}, James M. and {Swesty}, Douglas F.},
        title = "{A generalized equation of state for hot, dense matter}",
      journal = {\nphysa},
         year = 1991,
        month = dec,
       volume = {535},
       number = {2},
        pages = {331-376},
          doi = {10.1016/0375-9474(91)90452-C},
       adsurl = {https://ui.adsabs.harvard.edu/abs/1991NuPhA.535..331L}
}

@ARTICLE{2006A&A...445..273M,
       author = {{Marek}, A. and {Dimmelmeier}, H. and {Janka}, H. -Th. and {M{\"u}ller}, E. and {Buras}, R.},
        title = "{Exploring the relativistic regime with Newtonian hydrodynamics: an improved effective gravitational potential for supernova simulations}",
      journal = {\aap},
         year = 2006,
        month = jan,
       volume = {445},
       number = {1},
        pages = {273-289},
          doi = {10.1051/0004-6361:20052840},
archivePrefix = {arXiv},
       eprint = {astro-ph/0502161},
 primaryClass = {astro-ph},
       adsurl = {https://ui.adsabs.harvard.edu/abs/2006A&A...445..273M}
}

@ARTICLE{2017PhRvL.119p1101A,
       author = {{Abbott}, B.~P. and {Abbott}, R. and {Abbott}, T.~D. and {Acernese}, F. and {Ackley}, K. and {Adams}, C. and {Adams}, T. and {Addesso}, P. and {Adhikari}, R.~X. and {Adya}, V.~B. and {Affeldt}, C. and {Afrough}, M. and {Agarwal}, B. and {Agathos}, M. and {Agatsuma}, K. and {Aggarwal}, N. and {Aguiar}, O.~D. and {Aiello}, L. and {Ain}, A. and {Ajith}, P. and {Allen}, B. and {Allen}, G. and {Allocca}, A. and {Altin}, P.~A. and {Amato}, A. and {Ananyeva}, A. and {Anderson}, S.~B. and {Anderson}, W.~G. and {Angelova}, S.~V. and {Antier}, S. and {Appert}, S. and {Arai}, K. and {Araya}, M.~C. and {Areeda}, J.~S. and {Arnaud}, N. and {Arun}, K.~G. and {Ascenzi}, S. and {Ashton}, G. and {Ast}, M. and {Aston}, S.~M. and {Astone}, P. and {Atallah}, D.~V. and {Aufmuth}, P. and {Aulbert}, C. and {AultONeal}, K. and {Austin}, C. and {Avila-Alvarez}, A. and {Babak}, S. and {Bacon}, P. and {Bader}, M.~K.~M. and {Bae}, S. and {Bailes}, M. and {Baker}, P.~T. and {Baldaccini}, F. and {Ballardin}, G. and {Ballmer}, S.~W. and {Banagiri}, S. and {Barayoga}, J.~C. and {Barclay}, S.~E. and {Barish}, B.~C. and {Barker}, D. and {Barkett}, K. and {Barone}, F. and {Barr}, B. and {Barsotti}, L. and {Barsuglia}, M. and {Barta}, D. and {Barthelmy}, S.~D. and {Bartlett}, J. and {Bartos}, I. and {Bassiri}, R. and {Basti}, A. and {Batch}, J.~C. and {Bawaj}, M. and {Bayley}, J.~C. and {Bazzan}, M. and {B{\'e}csy}, B. and {Beer}, C. and {Bejger}, M. and {Belahcene}, I. and {Bell}, A.~S. and {Berger}, B.~K. and {Bergmann}, G. and {Bernuzzi}, S. and {Bero}, J.~J. and {Berry}, C.~P.~L. and {Bersanetti}, D. and {Bertolini}, A. and {Betzwieser}, J. and {Bhagwat}, S. and {Bhandare}, R. and {Bilenko}, I.~A. and {Billingsley}, G. and {Billman}, C.~R. and {Birch}, J. and {Birney}, R. and {Birnholtz}, O. and {Biscans}, S. and {Biscoveanu}, S. and {Bisht}, A. and {Bitossi}, M. and {Biwer}, C. and {Bizouard}, M.~A. and {Blackburn}, J.~K. and {Blackman}, J. and {Blair}, C.~D. and {Blair}, D.~G. and {Blair}, R.~M. and {Bloemen}, S. and {Bock}, O. and {Bode}, N. and {Boer}, M. and {Bogaert}, G. and {Bohe}, A. and {Bondu}, F. and {Bonilla}, E. and {Bonnand}, R. and {Boom}, B.~A. and {Bork}, R. and {Boschi}, V. and {Bose}, S. and {Bossie}, K. and {Bouffanais}, Y. and {Bozzi}, A. and {Bradaschia}, C. and {Brady}, P.~R. and {Branchesi}, M. and {Brau}, J.~E. and {Briant}, T. and {Brillet}, A. and {Brinkmann}, M. and {Brisson}, V. and {Brockill}, P. and {Broida}, J.~E. and {Brooks}, A.~F. and {Brown}, D.~A. and {Brown}, D.~D. and {Brunett}, S. and {Buchanan}, C.~C. and {Buikema}, A. and {Bulik}, T. and {Bulten}, H.~J. and {Buonanno}, A. and {Buskulic}, D. and {Buy}, C. and {Byer}, R.~L. and {Cabero}, M. and {Cadonati}, L. and {Cagnoli}, G. and {Cahillane}, C. and {Calder{\'o}n Bustillo}, J. and {Callister}, T.~A. and {Calloni}, E. and {Camp}, J.~B. and {Canepa}, M. and {Canizares}, P. and {Cannon}, K.~C. and {Cao}, H. and {Cao}, J. and {Capano}, C.~D. and {Capocasa}, E. and {Carbognani}, F. and {Caride}, S. and {Carney}, M.~F. and {Carullo}, G. and {Casanueva Diaz}, J. and {Casentini}, C. and {Caudill}, S. and {Cavagli{\`a}}, M. and {Cavalier}, F. and {Cavalieri}, R. and {Cella}, G. and {Cepeda}, C.~B. and {Cerd{\'a}-Dur{\'a}n}, P. and {Cerretani}, G. and {Cesarini}, E. and {Chamberlin}, S.~J. and {Chan}, M. and {Chao}, S. and {Charlton}, P. and {Chase}, E. and {Chassande-Mottin}, E. and {Chatterjee}, D. and {Chatziioannou}, K. and {Cheeseboro}, B.~D. and {Chen}, H.~Y. and {Chen}, X. and {Chen}, Y. and {Cheng}, H. -P. and {Chia}, H. and {Chincarini}, A. and {Chiummo}, A. and {Chmiel}, T. and {Cho}, H.~S. and {Cho}, M. and {Chow}, J.~H. and {Christensen}, N. and {Chu}, Q. and {Chua}, A.~J.~K. and {Chua}, S.},
        title = "{GW170817: Observation of Gravitational Waves from a Binary Neutron Star Inspiral}",
      journal = {\prl},
         year = 2017,
        month = oct,
       volume = {119},
       number = {16},
          eid = {161101},
        pages = {161101},
          doi = {10.1103/PhysRevLett.119.161101},
archivePrefix = {arXiv},
       eprint = {1710.05832},
 primaryClass = {gr-qc},
       adsurl = {https://ui.adsabs.harvard.edu/abs/2017PhRvL.119p1101A}
}

@ARTICLE{2023ApJS..268...66R,
       author = {{Reichert}, M. and {Winteler}, C. and {Korobkin}, O. and {Arcones}, A. and {Bliss}, J. and {Eichler}, M. and {Frischknecht}, U. and {Fr{\"o}hlich}, C. and {Hirschi}, R. and {Jacobi}, M. and {Kuske}, J. and {Mart{\'\i}nez-Pinedo}, G. and {Martin}, D. and {Mocelj}, D. and {Rauscher}, T. and {Thielemann}, F. -K.},
        title = "{The Nuclear Reaction Network WinNet}",
      journal = {\apjs},
         year = 2023,
        month = oct,
       volume = {268},
       number = {2},
          eid = {66},
        pages = {66},
          doi = {10.3847/1538-4365/acf033},
archivePrefix = {arXiv},
       eprint = {2305.07048},
 primaryClass = {astro-ph.IM},
       adsurl = {https://ui.adsabs.harvard.edu/abs/2023ApJS..268...66R}
}

@ARTICLE{2015MNRAS.454.1238L,
       author = {{Leung}, S. -C. and {Chu}, M. -C. and {Lin}, L. -M.},
        title = "{A new hydrodynamics code for Type Ia supernovae}",
      journal = {\mnras},
         year = 2015,
        month = dec,
       volume = {454},
       number = {2},
        pages = {1238-1259},
          doi = {10.1093/mnras/stv1923},
archivePrefix = {arXiv},
       eprint = {1507.08549},
 primaryClass = {astro-ph.HE},
       adsurl = {https://ui.adsabs.harvard.edu/abs/2015MNRAS.454.1238L}
}

@ARTICLE{2017ApJ...842...13W,
       author = {{Wongwathanarat}, Annop and {Janka}, Hans-Thomas and {M{\"u}ller}, Ewald and {Pllumbi}, Else and {Wanajo}, Shinya},
        title = "{Production and Distribution of $^{44}$Ti and $^{56}$Ni in a Three-dimensional Supernova Model Resembling Cassiopeia A}",
      journal = {\apj},
         year = 2017,
        month = jun,
       volume = {842},
       number = {1},
          eid = {13},
        pages = {13},
          doi = {10.3847/1538-4357/aa72de},
archivePrefix = {arXiv},
       eprint = {1610.05643},
 primaryClass = {astro-ph.HE},
       adsurl = {https://ui.adsabs.harvard.edu/abs/2017ApJ...842...13W}
}

@ARTICLE{2021ApJ...921..113S,
       author = {{Sandoval}, Michael A. and {Hix}, W. Raphael and {Messer}, O.~E. Bronson and {Lentz}, Eric J. and {Harris}, J. Austin},
        title = "{Three-dimensional Core-collapse Supernova Simulations with 160 Isotopic Species Evolved to Shock Breakout}",
      journal = {\apj},
         year = 2021,
        month = nov,
       volume = {921},
       number = {2},
          eid = {113},
        pages = {113},
          doi = {10.3847/1538-4357/ac1d49},
archivePrefix = {arXiv},
       eprint = {2106.01389},
 primaryClass = {astro-ph.HE},
       adsurl = {https://ui.adsabs.harvard.edu/abs/2021ApJ...921..113S}
}

@ARTICLE{2020PhRvD.102l3015B,
       author = {{Betranhandy}, Aurore and {O'Connor}, Evan},
        title = "{Impact of neutrino pair-production rates in core-collapse supernovae}",
      journal = {\prd},
         year = 2020,
        month = dec,
       volume = {102},
       number = {12},
          eid = {123015},
        pages = {123015},
          doi = {10.1103/PhysRevD.102.123015},
archivePrefix = {arXiv},
       eprint = {2010.02261},
 primaryClass = {astro-ph.HE},
       adsurl = {https://ui.adsabs.harvard.edu/abs/2020PhRvD.102l3015B}
}

@ARTICLE{2025RNAAS...9..113B,
       author = {{Brady}, Ryan and {Zingale}, Michael},
        title = "{Numerical Treatment of Shock-induced Nuclear Burning in Double Detonation Type Ia Supernovae}",
      journal = {Research Notes of the American Astronomical Society},
         year = 2025,
        month = may,
       volume = {9},
       number = {5},
          eid = {113},
        pages = {113},
          doi = {10.3847/2515-5172/add686},
archivePrefix = {arXiv},
       eprint = {2505.07918},
 primaryClass = {astro-ph.SR},
       adsurl = {https://ui.adsabs.harvard.edu/abs/2025RNAAS...9..113B}
}

@ARTICLE{2014ApJ...782...12P,
       author = {{Papatheodore}, Thomas L. and {Messer}, O.~E. Bronson},
        title = "{On Numerical Considerations for Modeling Reactive Astrophysical Shocks}",
      journal = {\apj},
         year = 2014,
        month = feb,
       volume = {782},
       number = {1},
          eid = {12},
        pages = {12},
          doi = {10.1088/0004-637X/782/1/12},
archivePrefix = {arXiv},
       eprint = {1312.5591},
 primaryClass = {astro-ph.SR},
       adsurl = {https://ui.adsabs.harvard.edu/abs/2014ApJ...782...12P}
}

@book{fryxell1989hydrodynamics,
  title={Hydrodynamics and nuclaer burning},
  author={Fryxell, B. and M{\"u}ller, E. and Arnett, D.},
  series={Max-Planck-Institut f{\"u}r Physik und Astrophysik M{\"u}nchen: MPA},
  url={https://books.google.com/books?id=4LQhtwAACAAJ},
  year={1989},
  publisher={Max-Planck-Inst. f{\"u}r Physik und Astrophysik}
}

@ARTICLE{2020A&A...635A.169G,
       author = {{Gronow}, Sabrina and {Collins}, Christine and {Ohlmann}, Sebastian T. and {Pakmor}, R{\"u}diger and {Kromer}, Markus and {Seitenzahl}, Ivo R. and {Sim}, Stuart A. and {R{\"o}pke}, Friedrich K.},
        title = "{SNe Ia from double detonations: Impact of core-shell mixing on the carbon ignition mechanism}",
      journal = {\aap},
         year = 2020,
        month = mar,
       volume = {635},
          eid = {A169},
        pages = {A169},
          doi = {10.1051/0004-6361/201936494},
archivePrefix = {arXiv},
       eprint = {2002.00981},
 primaryClass = {astro-ph.SR},
       adsurl = {https://ui.adsabs.harvard.edu/abs/2020A&A...635A.169G}
}

@ARTICLE{2015JCoPh.286..172C,
       author = {{Cavaglieri}, Daniele and {Bewley}, Thomas},
        title = "{Low-storage implicit/explicit Runge-Kutta schemes for the simulation of stiff high-dimensional ODE systems}",
      journal = {Journal of Computational Physics},
         year = 2015,
        month = apr,
       volume = {286},
        pages = {172-193},
          doi = {10.1016/j.jcp.2015.01.031},
       adsurl = {https://ui.adsabs.harvard.edu/abs/2015JCoPh.286..172C}
}

@ARTICLE{2025ApJ...978L..38C,
       author = {{Cheong}, Patrick Chi-Kit and {Pitik}, Tetyana and {Longo Micchi}, Lu{\'\i}s Felipe and {Radice}, David},
        title = "{Gamma-Ray Bursts and Kilonovae from the Accretion-induced Collapse of White Dwarfs}",
      journal = {\apjl},
         year = 2025,
        month = jan,
       volume = {978},
       number = {2},
          eid = {L38},
        pages = {L38},
          doi = {10.3847/2041-8213/ada1cc},
archivePrefix = {arXiv},
       eprint = {2410.10938},
 primaryClass = {astro-ph.HE},
       adsurl = {https://ui.adsabs.harvard.edu/abs/2025ApJ...978L..38C}
}

@ARTICLE{2024ApJ...966..150Z,
       author = {{Zingale}, Michael and {Chen}, Zhi and {Rasmussen}, Melissa and {Polin}, Abigail and {Katz}, Max and {Smith Clark}, Alexander and {Johnson}, Eric T.},
        title = "{Sensitivity of Simulations of Double-detonation Type Ia Supernovae to Integration Methodology}",
      journal = {\apj},
         year = 2024,
        month = may,
       volume = {966},
       number = {2},
          eid = {150},
        pages = {150},
          doi = {10.3847/1538-4357/ad3441},
archivePrefix = {arXiv},
       eprint = {2309.01802},
 primaryClass = {astro-ph.HE},
       adsurl = {https://ui.adsabs.harvard.edu/abs/2024ApJ...966..150Z}
}

@ARTICLE{2019ApJ...870...98U,
       author = {{Uchida}, Haruki and {Shibata}, Masaru and {Takahashi}, Koh and {Yoshida}, Takashi},
        title = "{Black Hole Formation and Explosion from Rapidly Rotating Very Massive Stars}",
      journal = {\apj},
         year = 2019,
        month = jan,
       volume = {870},
       number = {2},
          eid = {98},
        pages = {98},
          doi = {10.3847/1538-4357/aaf39e},
archivePrefix = {arXiv},
       eprint = {1809.10502},
 primaryClass = {astro-ph.HE},
       adsurl = {https://ui.adsabs.harvard.edu/abs/2019ApJ...870...98U}
}

@ARTICLE{pynucastro2,
doi = {10.3847/1538-4357/acbaff},
url = {https://dx.doi.org/10.3847/1538-4357/acbaff},
year = {2023},
month = {apr},
publisher = {The American Astronomical Society},
volume = {947},
number = {2},
pages = {65},
author = {Alexander I. Smith and Eric T. Johnson and Zhi Chen and
          Kiran Eiden and Donald E. Willcox and Brendan Boyd and
          Lyra Cao and Christopher J. DeGrendele and Michael Zingale},
title = {pynucastro: A Python Library for Nuclear Astrophysics},
journal = {The Astrophysical Journal}
}

@article{pynucastro,
   author = {{Willcox}, D.~E. and {Zingale}, M.},
    title = "{pynucastro: an interface to nuclear reaction rates and code generator for reaction network equations}",
  journal = {Journal of Open Source Software},
     year = 2018,
   volume = 3,
   number = 23,
   pages = 588,
     url = {https://doi.org/10.21105/joss.00588},
     doi = {10.21105/joss.00588},
 subject = {nuclear astrophysics}
}

@software{pynucastro_development_team_2025_17455462,
  author       = {pynucastro development team and
                  Boyd, Brendan and
                  Cao, Lyra and
                  Chen, Zhi and
                  Eiden, Kiran and
                  Glosser, Sam and
                  Johnson, Eric and
                  Li, Xinlong and
                  Rasmussen, Melissa and
                  Smith Clark, Alexander and
                  Willcox, Donald and
                  Zingale, Michael},
  title        = {pynucastro/pynucastro: pynucastro 2.8.0},
  month        = oct,
  year         = 2025,
  publisher    = {Zenodo},
  version      = {2.8.0},
  doi          = {10.5281/zenodo.17455462},
  url          = {https://doi.org/10.5281/zenodo.17455462},
  swhid        = {swh:1:dir:16d4c24a02eb286c41d26ef94b1c56ee489a8ef3
                   ;origin=https://doi.org/10.5281/zenodo.1202434;vis
                   it=swh:1:snp:24a21f6da32bde0b4ad08d5004b235cc44b0f
                   36b;anchor=swh:1:rel:ca5378926f65bfc0628397330f9d1
                   516fe9b8376;path=pynucastro-pynucastro-8141603
                  },
}

@ARTICLE{1984JCoPh..54..174C,
       author = {{Colella}, P. and {Woodward}, Paul R.},
        title = "{The Piecewise Parabolic Method (PPM) for Gas-Dynamical Simulations}",
      journal = {Journal of Computational Physics},
         year = 1984,
        month = sep,
       volume = {54},
        pages = {174-201},
          doi = {10.1016/0021-9991(84)90143-8},
       adsurl = {https://ui.adsabs.harvard.edu/abs/1984JCoPh..54..174C}
}

@ARTICLE{1982ApJ...258..696W,
       author = {{Wallace}, R.~K. and {Woosley}, S.~E. and {Weaver}, T.~A.},
        title = "{The thermonuclear model for X-ray transients}",
      journal = {\apj},
         year = 1982,
        month = jul,
       volume = {258},
        pages = {696-715},
          doi = {10.1086/160119},
       adsurl = {https://ui.adsabs.harvard.edu/abs/1982ApJ...258..696W}
}
\bibliographystyle{aasjournal}



\end{document}